**Transfer of Large-Scale Two-Dimensional Semiconductors: Challenges and Developments**

*Adam J. Watson, Wenbo Lu, Marcos H. D. Guimarães\* and Meike Stöhr\**

A. J. Watson, W. Lu, Prof. M. Guimaraes, Prof. M. A. Stöhr
Zernike Institute for Advanced Materials (ZIAM)
University of Groningen
Nijenborgh 4, 9747 AG, The Netherlands
Email: m.h.guimaraes@rug.nl and m.a.stohr@rug.nl

Abstract: Two-dimensional (2D) materials offer opportunities to explore both fundamental science and applications in the limit of atomic thickness. Beyond the prototypical case of graphene, other 2D materials have recently come to the fore. Of particular technological interest are 2D semiconductors, of which the family of materials known as the group-VI transition metal dichalcogenides (TMDs) has attracted much attention. The presence of a bandgap allows for the fabrication of high on-off ratio transistors and optoelectronic devices, as well as valley/spin polarized transport. The technique of chemical vapour deposition (CVD) has produced high-quality and contiguous wafer-scale 2D films, however, they often need to be transferred to arbitrary substrates for further investigation. In this Review, the various transfer techniques developed for transferring 2D films will be outlined and compared, with particular emphasis given to CVD-grown TMDs. Each technique suffers undesirable process-related drawbacks such as bubbles, residue or wrinkles, which can degrade device performance by for instance reducing electron mobility. This Review aims to address these problems and provide a systematic overview of key methods to characterize and improve the quality of the transferred films and heterostructures. With the maturing technological status of CVD-grown 2D materials, a robust transfer toolbox is vital.

1. **Introduction**

Ever since the discovery and isolation of graphene by Geim and Novoselov in 2004[1] using the technique of mechanical exfoliation (the 'Scotch tape' method), the field of two-dimensional (2D) materials has become one of the most intensively researched in condensed matter physics. 2D layered materials (2DLMs) offer opportunities to explore fundamental physics in the limit of atomic thickness, and have advantages over existing materials with regards to technological applications.[2] Graphene, the prototypical 2D material, has numerous interesting properties, including a high mobility, transparency, tensile strength, etc.[3] However, lacking a bandgap[4], it is limited in its applications, for instance in optoelectronics and for conventional field-effect transistors (FETs).[5] Hence, other 2D materials have been investigated. Transition metal dichalcogenides (TMDs) are a class of materials with a rich catalogue of novel properties, many of which go beyond those of graphene. Their general formula is given as $MX_2$, where M is a transition metal atom, and X is a chalcogen atom (usually S, Se or Te). The group-VIB TMDs (e.g. $MoS_2$ and $WSe_2$) represent the most extensively studied in the monolayer (ML) limit. The exciting technological potential has been realized in the demonstration of atomically thin FETs[6–9], tunable photovoltaic or light emitting devices for optoelectronic applications[10–12], as well as more exotic devices based on spin-valley coupling.[13]

Initial research on 2D TMDs relied on mechanical exfoliation from a bulk crystal. However, this method yields unpredictable flake thickness and domain sizes, which are usually on the order of a few microns. Moreover, the method is relatively time consuming. To meet the demands placed upon TMDs with respect to their technological applications, two conditions must be met. Firstly, scalable production methods are required. High quality 2D TMDs, of wafer scale, are needed to produce integrated circuits (ICs), compatible with existing industrial fabrication methods. Recently, the technique of chemical vapor deposition (CVD) has been used to successfully grow large area TMD films (up to centimeter scale) with high uniformity, in a cost effective manner.[14–18] Samples made using this method have shown properties on par with, or even surpassing, those of exfoliated films.[19] The second condition is flexibility over substrate choice. This remains a challenge for the CVD method, as the target substrates for TMD-based devices may not be able to withstand the high-temperature environment produced during the CVD growth process.[20] Furthermore, it is also desirable to fabricate heterostructures from individual TMD films, with customizable stacking order. This requires a systematic methodology for transferring large-scale TMD films from their growth substrate onto a target substrate, while maintaining the intrinsic structural and physicochemical properties that make 2D TMDs so appealing. Hence, any successful transfer method must allow for a uniform separation of the film from its growth substrate, and also maintain the structural integrity of the film during the transfer steps.

Many transfer techniques were developed initially to transfer mechanically exfoliated flakes of graphene. One of the most common methods uses polymethyl methacrylate (PMMA) as a support layer to transfer the exfoliated flake to the desired target substrate.[21,22] Such a method has been successful at transferring exfoliated flakes to diverse substrates. However, modifications to this method are required for CVD-grown 2DLMs. Substrates used in CVD (such as $SiO_2$/Si or mica) do not have a water-soluble layer commonly used in the standard PMMA method for exfoliated flakes, requiring other methods to remove the 2DLM from the growth substrate. Often this entails harsh chemical etchants such as KOH, which can damage the 2DLM. Furthermore, because of the size of the film being transferred (up to centimeter scale), maintaining the structural integrity of the film is of paramount importance to ensure a uniform transfer. Thus, mechanical supports take on a more critical role. Polymers, including PMMA, fulfil this role. However, the problems associated with using these materials, such as cracks, wrinkles or polymer residue, have led to a search for other materials to serve as supports, and some methods forgo the use of any support entirely. **Figure 1** illustrates some of these problems.

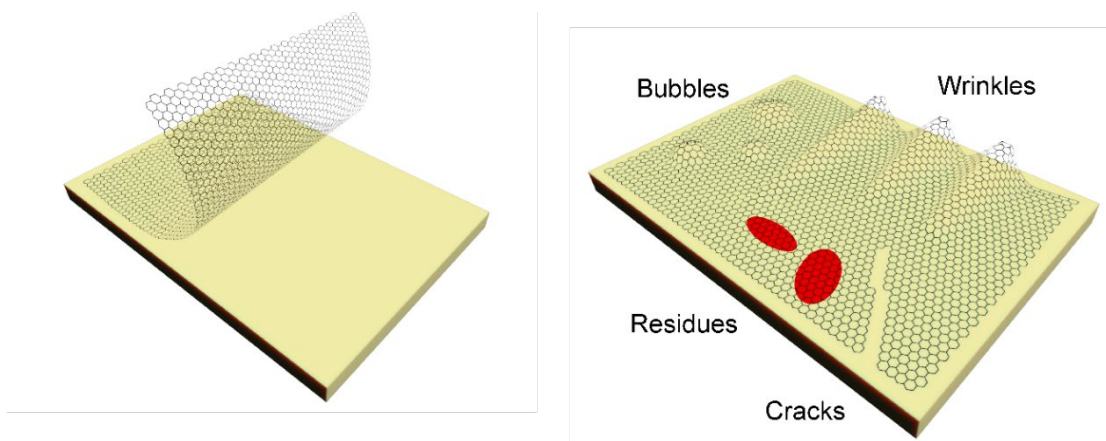

**Figure 1.** Schematic illustration of the transfer of a 2DLM onto an arbitrary substrate. The right image indicates the kinds of issues that are encountered from the transfer procedure, including cracks, residues from the supports, trapped bubbles, and wrinkles.

The purpose of this Review is to provide an overview of the transfer methodologies currently used to transfer CVD-grown TMDs, and to offer a means to quantify and potentially solve their process-related drawbacks. Crucially, the Review is written with an eye to industrial applications. This will provide a grounding for fledgling researchers who are starting their work on CVD-grown 2D materials and the fabrication of van der Waals heterostructures. To this end, Section 2 will begin with a scheme to categorize the various transfer methods. The similarities and, perhaps more crucially, the differences between graphene and TMDs will be outlined. This is important as many of the transfer methods that work with graphene may not work identically with TMDs, owing to the different structural make-up of each material type. The various transfer techniques will be outlined, and the section will conclude with a critical comparison between each method. In Section 3, the process related drawbacks often encountered in transferring CVD TMDs will be discussed. Problems such as polymer residue, cracks or wrinkles (to name a few) are encountered routinely in CVD-based transfer. These often degrade the properties of the film and fabricated devices. However, a comprehensive investigation has not yet been done on these problems, which need to be solved if TMDs are to be industrially applicable. To this end, an overview of the various problems with transfer will be given, as well as a description of the techniques to characterize them quantitatively. Finally, a conclusion will draw together the various threads of the Review to provide a commentary on the current state of transfer techniques for large area CVD-grown TMDs, and an outlook on future developments.

**2. Transfer methods for 2D TMDs**

In this section, a categorization scheme for the transfer of 2D TMDs is introduced. Many of these techniques, originally developed for the deterministic transfer of exfoliated flakes, are now finding application for CVD-grown 2DLMs, with some modifications. These modifications are needed due to the larger size of CVD-grown 2DLMs. Issues relating to surface energetics, film quality and uniformity, and structural supports are more prominent, as the spatial variation of forces can lead to film breakage or other undesirable effects. In addition to the mechanical differences between exfoliated and CVD materials transfer, the transfer of large-scale 2DLMs is highly relevant for technological applications. It is therefore instructive to provide an overview of the various large-scale transfer techniques employed so far, including a discussion of their advantages and disadvantages, and also describe how they have been used for CVD-grown TMD 2DLMs. Traditionally, transfer techniques are classified into 'wet' and 'dry' categories. Dry transfer involves no direct contact of the 2DLM with water or chemicals during the main transfer step. Wet techniques involve the delamination of the 2D films from their original substrates using water, or chemicals in the liquid phase. Within these categories, a more convenient delineation can be made into methods that use supports (such as polymers), and those that do not. This is because many transfer methods use a mixture of wet and dry techniques, resulting in some ambiguity when using the traditional classification. In contrast, supporting layers (or lack thereof) provide a clearer means of distinguishing between the various methods. Each of these methods will be described in detail below. A comparative overview of some of the key mechanical properties between TMDs and supports is given in **Table 1**.

**Table 1.** Mechanical properties of TMDs and their supports.

| Materials | Surface Energy [mJ m$^{-2}$] | Young's Modulus [Gpa] | TEC [× 10$^{-6}$ K$^{-1}$] |
|---|---|---|---|
| MoS$_2$ | 35-48.3[23,24] | 270[25] | 7.6[26] |
| MoSe$_2$ | Unav. | 177[27] | 7.24[28] |
| WS$_2$ | 39[24] | 272[29] | 10.3[30] |
| WSe$_2$ | Unav. | 167[31] | 14.5[32] |
| PMMA | 41[33] | $8 \times 10^{-3}$ [34] | 180[35] |
| PDMS | 19.8[33] | $3.6\text{-}8.7 \times 10^{-4}$ [36] | 906[35] |
| PVP | 48-63[37] | 0.12[38] | Unav. |
| PS | 40.7[33] | 3.5[39] | 200[35] |
| PVA | 36.5[33] | $1.6 \times 10^{-2}$ [40] | Unav. |
| CA | 35.04[41] | 2[42] | 73[42] |
| Cu | 1650[43] | 100[44] | 16.7[45] |

## 2.1 TMD transfer with a support layer

One of the first methods developed to transfer 2DLMs involved using a supporting layer on top to better control the strain and forces during transfer. Polymers are the material of choice due to their flexibility, mechanical strength and ability to form a uniform contact with the 2DLM, but other supports (such as thin metallic films) have been used as well. The majority of TMD transfer techniques developed so far find their origins in those developed for graphene transfer [21,22,46–48], but the underlying principles are similar. This is mainly a result of the fact that both materials are van der Waals (vdW) materials, meaning the surface energetics are similar. The surface energy of a material can be described by Young's equation, written as

$$\sigma_{sg} = \sigma_{sl} + \sigma_{lg} \cos \theta \qquad (1)$$

where $\theta$ is the contact angle between the liquid and solid, $\sigma_{lg}$ is the surface tension of the liquid, $\sigma_{sl}$ is the interfacial tension between the liquid and solid, and $\sigma_{sg}$ is the surface free energy of the solid in units of J m$^{-2}$. A schematic outlining the various terms is shown in **Figure 2**. In general, a hydrophobic surface will give a contact angle of $\geq 90°$, resulting in a low surface energy, whereas a hydrophilic surface will give a contact angle of $< 90°$, giving a higher surface energy.

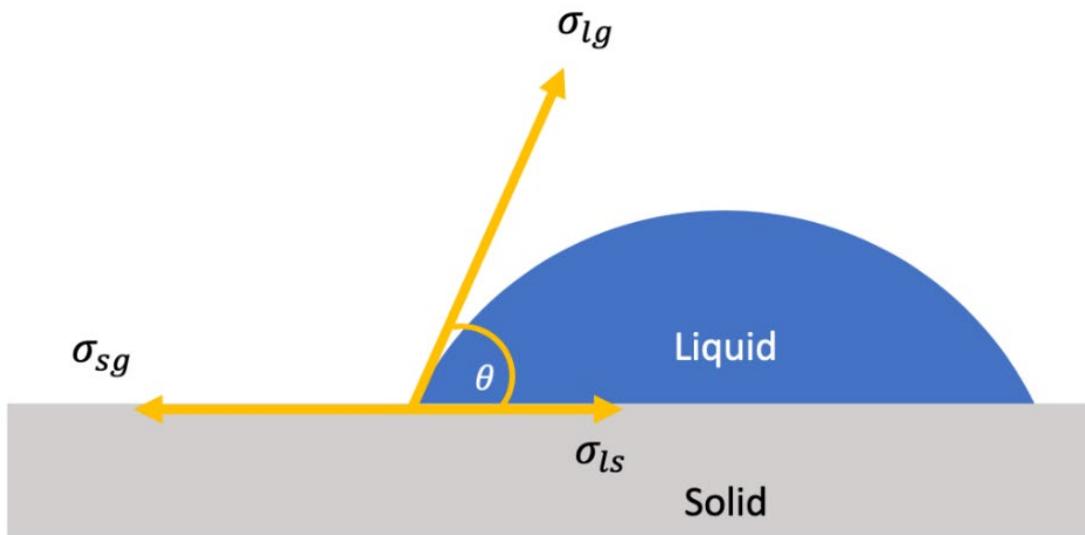

**Figure 2.** A schematic diagram of the components of a 3-phase system relevant to determining the surface energy of a material. $\sigma_{sg}$, $\sigma_{ls}$ and $\sigma_{lg}$ represent the surface tension (in J m$^{-2}$) of the solid-gas interface, the liquid-solid interface and the liquid-gas interface, respectively. As the angle $\theta$ approaches 90°, $\sigma_{sg}$ is lowered.

As illustrated by the equation (1), the proper choice of polymer support is affected by the surface energy of the polymer, the growth substrate and the destination substrate, ultimately determining the quality of the transferred film.[49] A lower surface energy corresponds to a lower adhesion force[50], meaning that polymers with lower surface energy will be more easily removed with minimum damage or residues. On the other hand, the surface energy of the target substrate must be larger than that of the polymer, to ensure proper adhesion of the transferred film to the new substrate. Thus, care must be given to substrate and polymer choice.

Both graphene and TMDs offer a unique combination of mechanical properties, such as a high in plane stiffness and strength, as well as a low bending modulus. But despite the *prima facie* similarities between the two material types, there are some notable differences in their mechanical properties. For example, the Young's modulus of $MoS_2$ (130 N m$^{-1}$)[51] is less than half that of graphene (340 N m$^{-1}$)[52]. On the other hand, $MoS_2$ has a bending modulus of 9.61 eV[53], which is about 7 times larger than that of graphene (1.4 eV)[54]. This is due to the trilayer atomic structure of $MoS_2$, resulting in more interaction terms in the bending energy calculation, which restricts the bending motion. This has implications for transfer, as it means TMDs do not buckle as readily under external compression, which is advantageous compared to graphene. One of the most striking differences, however, is in the thermal expansion coefficient (TEC) of both materials. For TMDs, the TEC is positive. $MoS_2$, for instance, has a value of ~$7.6 \times 10^{-6}$ K$^{-1}$,[26] and for $WS_2$ it is ~$10 \times 10^{-6}$ K$^{-1}$.[30] For graphene, however, it is negative, with an average value of $-3.75 \times 10^{-6}$ K$^{-1}$.[55] Since the TEC of most polymers (and all metals) are positive, we would expect significant strain during any heating step, leading to wrinkles and cracks when using such supports for graphene transfer[56,57], and much less so for TMDs. Below, we look at some of the main methods of transferring TMD films using mechanical supports.

*2.1.1 Polymethyl methacrylate (PMMA)-assisted transfer*

First used as a support layer for the transfer of mechanically exfoliated flakes of graphene[22,58], the PMMA-assisted method was thereafter applied to CVD-grown graphene[59,60], and then to CVD-grown $MoS_2$ a few years later[61]. The standard procedure, for the case of CVD-grown $MoS_2$ on a $SiO_2$/Si substrate, is as follows[46]. (1) PMMA is spin-cast on top of $MoS_2$ followed by baking at 100 °C for 10 min. (2) The PMMA/$MoS_2$/$SiO_2$/Si stack is then placed in a hot NaOH solution (100 °C for 30 min) to etch the $SiO_2$ layer and detach the PMMA/$MoS_2$ from the substrate. (3) The PMMA/$MoS_2$ is rinsed with deionized (DI) water to remove remaining etchant and residues. (4) A target substrate is used to fish out the PMMA/$MoS_2$ film and then dried on a hot plate (100 °C for 10 min). (5) Finally, the PMMA is removed with acetone and isopropyl alcohol (IPA), and rinsed with DI water and chloroform. Each step can be adjusted accordingly. For instance, instead of NaOH, another hot base solution of for example KOH can be used to etch the $SiO_2$ layer.[62]

The use of strong chemicals to etch the growth substrate means it cannot be reused, making the process relatively expensive, and perhaps prohibiting its use in industrial applications. Therefore, a different method was developed to detach the CVD-grown film from the growth substrate without wet etching processes, whilst keeping the PMMA support. Initially developed to delaminate graphene[63,64], the so-called bubbling method uses bubble intercalation to weaken the adhesion between the 2D film and growth substrate. This method was used in a comparative study by Yun et al. in transferring centimeter scale monolayer CVD-grown $WS_2$ on Au foil[65]. For the bubbling method, the $WS_2$/Au assembly was immersed in a NaCl solution and used as a cathode. Hydrogen bubbles were produced between the $WS_2$ film and Au surface according to the reaction $2H_2O(l) + 2e^- \rightarrow H_2(gas) + 2OH^-(aq)$, separating the PMMA/$WS_2$ layer from the Au foil and allowing it to be reused. The process is outlined in **Figure 3**(a).

Another important development in the etching-free transfer process involves making use of a water-soluble sacrificial layer. Zhang et al. used a novel CVD method to grow $MoS_2$ flakes on top of a crystalline layer of $NaS_x$ and NaCl, on a $SiO_2$/Si substrate[66]. Precursors of both NaCl and $MoO_3$ in the ratio of 9:1, chosen due to their similar melting temperatures (795 °C and 801 °C respectively), are brought into the vapor phase. By exploiting the difference in dissociation energies of both the Na-O and Mo-O bonds, a layer of $NaO_x$ is formed first on the $SiO_2$/Si substrate and then vulcanized by sulfur vapor to form $NaS_x$, which is highly water soluble. Thereafter, $MoS_2$ clusters form on top. The transfer step involves spin-casting PMMA on the $MoS_2$ wafer, and then placing it in a container with a gradual addition of DI water. The $NaCl/NaS_x$ layer is dissolved and the $PMMA/MoS_2$ layer is transferred to a new substrate. PMMA removal is carried out via an acetone treatment.

Thermal release tape (TRT) provides another means by which TMD layers can be delaminated from their growth substrates, without the use of etchants or solvents.[16,67,68] Kang et al. made use of TRT and van der Waals stacking in vacuum to produce CVD-grown TMD heterostructures with clean interfaces.[16] PMMA was spin-cast first onto a monolayer TMD (called L0), and then TRT was used to peel it off the $SiO_2$/Si substrate. The TRT/L0 stack was then pressed onto another TMD layer which was subsequently peeled off. The process was repeated until the desired number of layers was reached, after which the layers were placed onto a target substrate. The TRT was removed by applying heat and the PMMA, which was on the top layer only, was removed via acetone. In this method, although PMMA was used as a support in the initial step, the subsequent stacking of the monolayers was done via the van der Waals interaction. This process is outlined in Figure 3(b).

The predominance of the PMMA-assisted method in the initial transfer of CVD-grown TMDs is largely a result of its use in transferring graphene films, where it serves as a robust supporting layer with good flexibility and adhesive contact.[69] A tried-and-tested methodology was developed which could simply be duplicated for use in transferring exfoliated films and, thereafter, larger-area TMD films for initial characterization. However, the harsh chemicals used to remove the PMMA, usually at elevated temperatures, results in damage to the TMD layer. Moreover, the method invariably leaves a residue which is hard to remove with post-transfer cleaning procedures such as ultra-high vacuum annealing.[67,70]

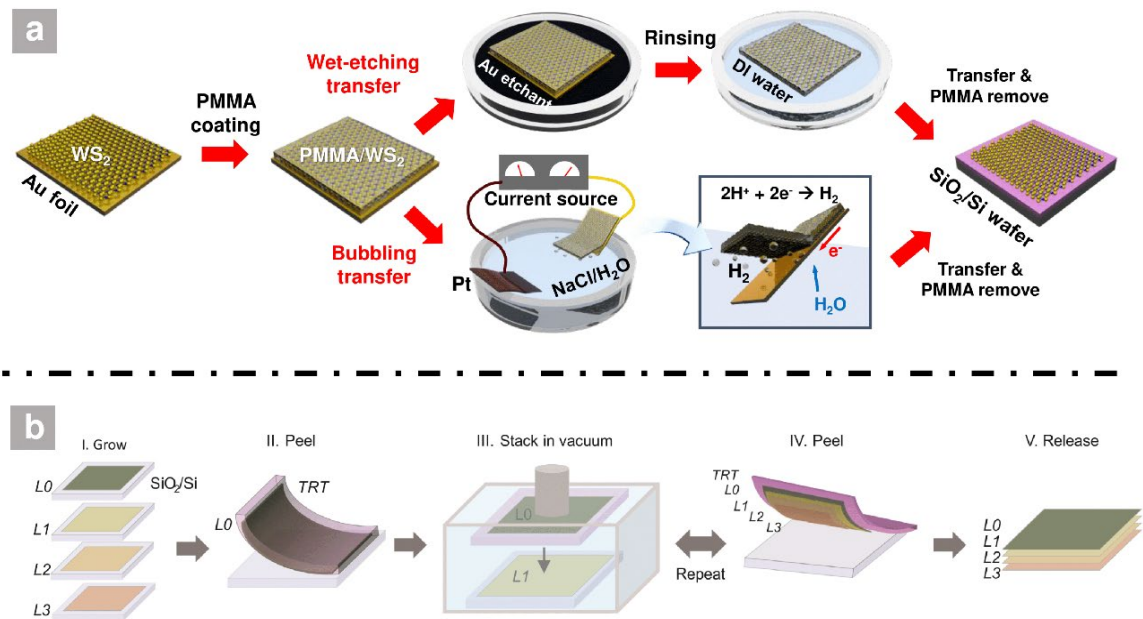

**Figure 3.** (a) Schematic illustrations of both the bubbling transfer and wet chemical etching techniques used in transferring $WS_2$ from Au foil to a $SiO_2$/Si substrate. For the bubbling method, the PMMA/$WS_2$/Au assembly is immersed in a NaCl solution, where the Au foil acts as a negatively charged cathode. Hydrogen bubbles are generated between the $WS_2$ film and Au foil, creating a pressure which delaminates the $WS_2$ film. In the final step, the PMMA is removed. The advantage of this method compared to the wet-etching method is that the Au foil can be reused. Reproduced with permission.[65] Copyright 2015, American Chemical Society. (b) The TRT assisted stacking process, carried out in vacuum. (I) First the individual layer is grown on $SiO_2$/Si substrates via CVD. (II) TRT is placed onto the PMMA-coated first layer (L0), and peeled off the growth substrate. (III) A stacking procedure is done in vacuum, in which L0 is used to pick up the next layer (L1), and the process can be repeated to achieve the desired number of layers in the stack. (IV) TRT is used to peel the completed stack off the last growth substrate. (V) The assembly is placed on a target substrate, and the TRT is released via heating. Reproduced with permission.[16] Copyright 2017, Springer Nature.

*2.1.2 Polydimethylsiloxane (PDMS)-assisted transfer*

Polydimethylsiloxane (PDMS) is a widely used organic polymer that has found application in the transfer of 2DLMs, due to its hydrophobicity, transparency, high flexibility and low surface energy.[71,72] In particular, the lower surface energy of PDMS (~19-21 mJ m$^{-2}$)[73] compared to that of common target substrates such as SiO$_2$/Si (57 mJ m$^{-2}$)[74] means that TMD layers can be detached from their PDMS supports with relative ease.[75] This means that the final step of removing the polymer layer with wet chemicals is not necessary, in principle resulting in a cleaner transfer. The use of PDMS in the transfer of CVD-grown TMD layers has been done[19,29,76–81], with notable variations in methodology. For instance, in order to increase the adhesion force between PDMS and the MoS$_2$ film, Kang et al. used hydrophilic dimethyl sulfoxide (DMSO) molecules in a DI water solution that were vaporized onto the PDMS surface at 270 °C, in order to increase the surface energy and therefore the adhesion force.[79] This meant the PDMS mold could pick up the whole MoS$_2$ film from the SiO$_2$/Si substrate. When the PDMS/MoS$_2$ was brought into contact with the target substrate at 70 °C, the standard adhesion force of PDMS was restored and the MoS$_2$ could successfully detach from the polymer stamp. This method, shown schematically in **Figure 4**(a), is an all-dry transfer process involving no wet chemical or etching steps. This has obvious advantages in terms of both transfer speed and resulting film cleanliness.

In an alternative method[19], Jia et al. took advantage of the hydrophobic PDMS stamp and the hydrophilic SiO$_2$/Si substrate to delaminate CVD MoS$_2$ using DI water droplets. This is schematically illustrated in Figure 4(b). The MoS$_2$/substrate was pressed onto the PDMS stamp attached to a glass slide. Water droplets injected at the edge of the stamp penetrated between the MoS$_2$ and substrate, detaching the MoS$_2$ film reliably. The PDMS stamp was subsequently removed by peeling. The advantage of this method over the DMSO-mediated method is the lack of a heating step, which can lead to structural damage.

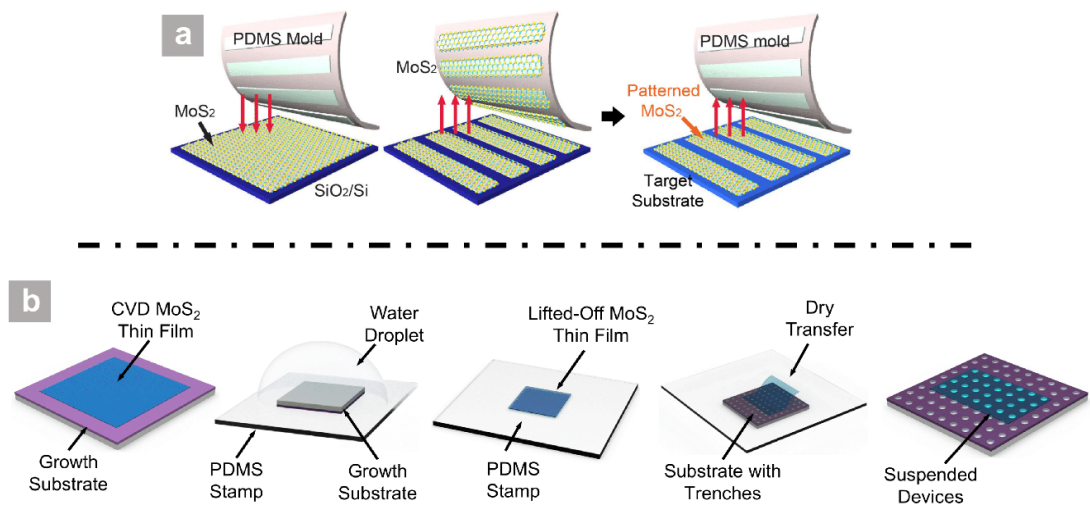

**Figure 4.** (a) Schematic illustration of the all-dry transfer method utilizing a PDMS-support layer to transfer CVD-grown MoS$_2$. A PDMS mold is brought into contact with the MoS$_2$ film on a SiO$_2$ substrate (left), and is then peeled away, removing strips of MoS$_2$ (middle). Finally, the MoS$_2$/PDMS strip assembly is brought into contact with a target substrate, allowing the MoS$_2$ layers to adhere to the surface before removing the PDMS mold (right). Reproduced with permission.[79] Copyright 2017, Elsevier. (b) Schematic illustration of PDMS transfer using both water-delamination and dry transfer methods. MoS$_2$ is grown using CVD (left). A PDMS stamp is brought into contact with the MoS$_2$ film and a water droplet is introduced from the side (second from left), leading to water intercalation and separating the MoS$_2$ film from the substrate (middle). The PDMS/MoS$_2$ assembly is brought into contact with a substrate with pre-patterned circular microtrench arrays (second from right), before the PDMS is removed by mechanical peeling leaving the film on the target substrate (right). Reproduced with permission.[19] Copyright 2016, Royal Society of Chemistry.

A notable modification to the PDMS-supported transfer procedure is to introduce polyvinyl alcohol (PVA) in between the PDMS and TMD film. This intermediate layer was introduced due to the relatively poor adhesion between PDMS and 2D materials.[82] Rather than using PDMS as the direct contact polymer with the 2DLM, PVA is attached to the PDMS and is used as the direct support. The PDMS serves as a secondary supporting layer for the PVA/2DLM assembly, and is attached to the glass slide. The PDMS/PVA procedure was carried out by Cao et al. to transfer a whole film of CVD-grown WSe$_2$ onto a SiO$_2$/Si substrate with pre-patterned electrodes.[83] With a larger PVA film, the authors predict that larger area films can be transferred. This scalability is appealing for applications. The use of PVA as a support has notable advantages, specifically in its water-solubility, as well as its good adhesion to 2DLMs. However, its use as a standalone support layer is hindered by its low viscoelastic properties[84], meaning that it does not provide a strong enough support to enable a uniform transfer. Hence, a secondary supporting layer is required. This can introduce additional complexity to the transfer process.

As mentioned above, the low surface energy of PDMS relative to various substrates can be problematic, particularly for detaching the film from growth substrates such as SiO$_2$/Si. Modification to the PDMS surface energy[79], or water intercalation[19] is required to assist the PDMS in delaminating the film. However, the low surface energy is an advantage in removing the PDMS from the TMD after it has been transferred. In addition, the presence of uncrosslinked oligomers (up to 5% depending on the curing time[85]) can remain on the surface of the TMD after transfer, causing contamination.[86] Hence, further treatment to fully remove this residue is required.

*2.1.3 Polystyrene (PS)-assisted transfer*

The well-known hydrophobic polymer polystyrene (PS) has also found application in the transfer of CVD-grown TMD layers.[48,87–93] For instance, Gurarslan et al. made use of a surface energy-assisted process to delaminate MoS$_2$.[92] A thin layer of PS was spin-cast onto an MoS$_2$ monolayer grown on a sapphire substrate, which was chosen because the (0001) plane of c-sapphire is hexagonal, thus matching the lattice symmetry of many TMDs. Making use of the different surface energies between film and substrate, the hydrophobic MoS$_2$ layer is delaminated from its hydrophilic growth substrate. The procedure is illustrated in **Figure 5**. First, a layer of PS was spin-cast on the as-grown MoS$_2$ film. A droplet of water was then added. The water intercalation was not automatic because of the strong film-substrate interaction, and required gently poking the side of the sample with a sharp object. Once started, the process was very fast, and the PS/MoS$_2$ assembly could be removed and dried with a paper towel. After the PS/MoS$_2$ stack was transferred to a SiO$_2$/Si substrate, the PS was removed by toluene.

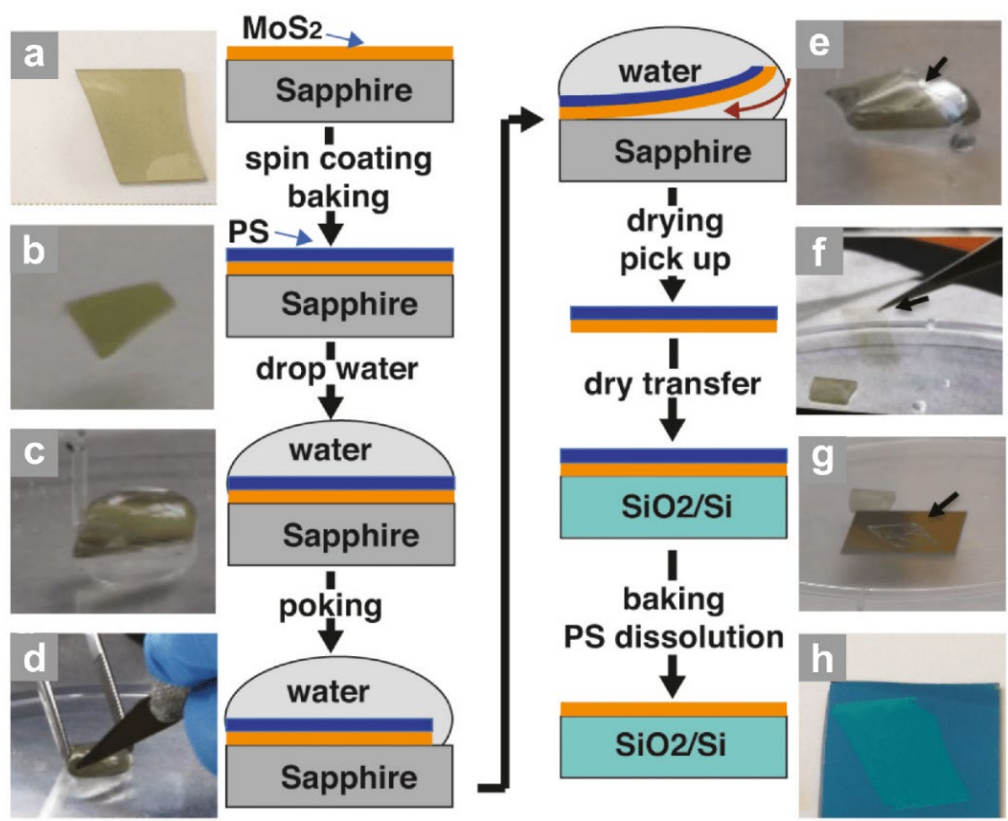

**Figure 5**. Schematic illustration of the PS-assisted transfer of MoS$_2$ on a sapphire substrate, with corresponding images of each step (a-h). The as-grown MoS$_2$ film (a) covered in a spin-coated layer of PS (b), after which a water droplet is added on top (c). By poking at the edge of the PS/MoS$_2$ assembly, the water can intercalate between the MoS$_2$ and sapphire substrate (d), eventually delaminating the PS/MoS$_2$ assembly (e). The PS/MoS$_2$ is lifted off and dried (f), and then placed onto the target SiO$_2$/Si substrate (g). The PS is dissolved by baking as a final step (h). Reproduced with permission.[92] Copyright 2014, American Chemical Society.

Xu et al. used a similar method to delaminate CVD-grown WS$_2$ from a sapphire substrate.[48] To improve the speed of delamination, the sample was pre-etched in NaOH solution for a number of minutes. Thereafter the sample was immersed in water to delaminate the PS/WS$_2$ from the sapphire. It was found that etching for 5 minutes resulted in WS$_2$ delamination within 30 seconds, whilst after 10 minutes of etching the delamination occurred instantaneously. However, for the latter case substantial damage to the sapphire substrate was incurred. Nevertheless, such a short etching time represents a substantial improvement over the commonly used etching times (typically 30-60 minutes) at elevated temperatures of up to 100 °C. The method represents an improvement over that described in ref [92] in two important respects. Firstly, the thickness of the PS film was made very thin (~100 nm), to avoid any residual stress obtained in thicker PS films that caused breaking of the MoS$_2$ flakes observed in ref[92]. Secondly, a controlled delamination process was employed, in which the sample was lowered into the water at a delamination rate of 0.3 cm$^2$ s$^{-1}$. Combined, a more uniform transfer of WS$_2$ was achieved.

PS has a number of advantages over the traditional PMMA-assisted method. For instance, PS has a larger Young's modulus (3.5 GPa)[39] than PMMA (8 MPa)[34]. This means it provides a more robust support to the TMD films, preventing wrinkling. Furthermore, the aromatic structure of PS allows for a wider range of solvents, such as tetrahydrofuran (THF). It was further found that the solubility of PS in THF was greater than PMMA in acetone.[94] However, PS is more brittle than PMMA, hindering its use in larger scale transfer. To solve this issue, a thinner PS film can be made (as in ref[48]). Alternatively, a modified form of PS can be used, in which the molecule 4,4'-diisopropylbiphenyl (DIPB) is mixed with PS to widen the distance between the polymer chains, making it softer and more mechanically flexible.[94,95]

*2.1.4 Other polymer-assisted transfer*

In addition to the above, other less common polymer supports have been used for transferring TMD layers, and ultimately expand the repertoire of supports for transferring CVD-grown TMDs. For this reason, they deserve to be highlighted in this review. Citing the well-known problems of using PMMA-based transfer methodologies, specifically polymer residues which degrade performance (see for example refs [22,70,96-98]), Zhang et al. used cellulose acetate (CA) as a support layer to transfer CVD-grown TMDs onto a $SiO_2$/Si substrate.[99] To avoid the problems associated with using the conventional hot NaOH etching method to detach the film (such as cracks or wrinkles from bubbles), the authors used a combination of $NH_4F$ and HF (known as buffered oxide etch, or BOE) which works at room temperature. A schematic illustration of the transfer procedure is shown in **Figure 6**(a). Firstly, a layer of CA was spin-coated onto the as-grown TMD film. A BOE etch for a few minutes without heating removed the $SiO_2$ layer and detached the TMD film, which was then transferred after DI rinsing to a $SiO_2$/Si target substrate. The CA was removed via IPA and acetone baths. Among the advantages of using CA is that it can be easily dissolved in acetone, which should in principle lead to less residues. Furthermore, CA is inexpensive, non-toxic and biodegradable[100], making it a suitable environmentally-friendly support.

An alternative, scalable transfer method was used to transfer wafer-scale CVD-grown TMD films (~6 inches), making use of the roll-to-roll production method that was developed for graphene transfer.[101,102] Yang et al. adopted this method to transfer CVD-grown $MoS_2$ on a glass substrate using an ethylene vinyl acetate/polyethylene terephthalate (EVA/PET) plastic support.[103] The method is outlined in Figure 6(b). Firstly, an EVA/PET film was attached to the $MoS_2$/glass assembly via a hot lamination method using a lamination machine. The EVA/PET/$MoS_2$/glass assembly was then immersed in DI water for about 5 minutes to delaminate the $MoS_2$ from the glass substrate followed by a drying step using $N_2$. The novelty of this particular method is that the polymer support also serves as the transferred substrate, thus requiring no polymer-removal step and paving the way for large-scale batch production of flexible electronic components.

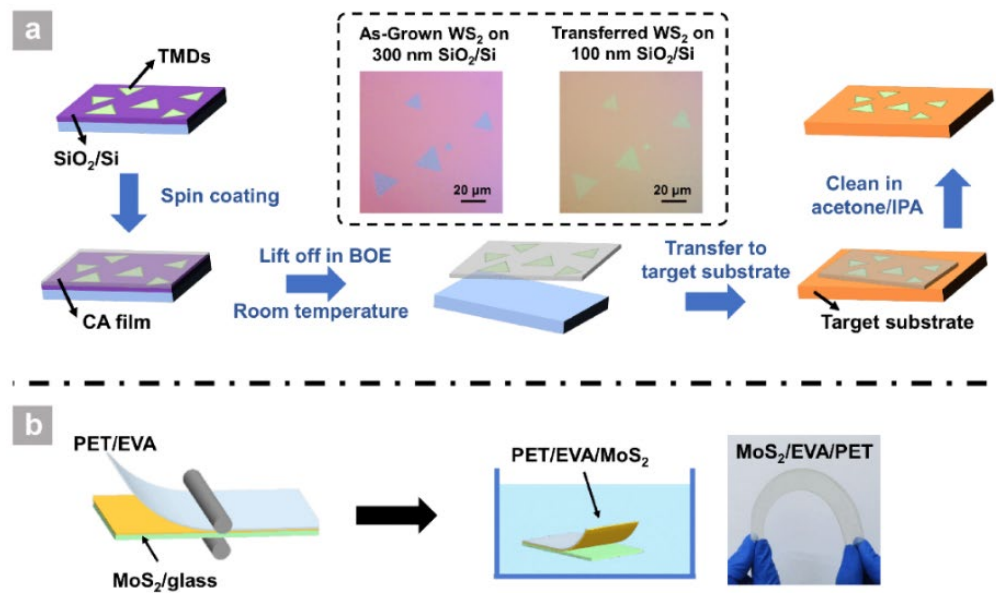

**Figure 6.** (a) Schematic illustration of the cellulose acetate (CA)-assisted transfer of CVD-grown TMD films. TMD films are grown on a SiO$_2$/Si substrate, followed by spin casting a layer of CA. The CA/TMD assembly is then placed on the surface of BOE to etch the SiO$_2$ layer and detach the TMD, and is then transferred to a target substrate and rinsed in acetone and IPA. Reproduced with permission.[99] Copyright 2019, American Chemical Society. (b) The green transfer method to transfer large-scale CVD-grown MoS$_2$ films on glass to an ethylene vinyl acetate/polyethylene terephthalate (PET/EVA) target, using the roll-to-roll production method. The MoS$_2$/glass is attached to the PET/EVA via a hot lamination method (left image). The MoS$_2$ is delaminated from the glass substrate via immersion into DI water (middle image). The transferred MoS$_2$ on EVA/PET is shown in the right image. Reproduced with permission.[103] Copyright 2018, Springer Nature.

In addition to its use with PDMS as described above in Section 2.1.2, Lu et al. used PVA in conjunction with polyvinylpyrrolidone (PVP) to form a water-soluble bilayer as a support.[104] PVP was used as the direct contact with the 2DLM, given its good adhesion and wetting properties. To improve the wettability, *N*-vinylpyrrolidone (NVP) was added to the PVP solution to match the surface energies of $MoS_2$ and $WS_2$. The PVA top layer serves as a structural support to reinforce the more flexible PVP. Notably, we see PVA being used again (see Section 2.1.2) in a bilayer polymer supporting structure, the difference being that here the bilayer is entirely water-soluble.

*2.1.5 Metal-assisted transfer*

Despite the efforts to minimize the problem of polymer residues, it remains an enduring issue in using polymer supports for transfer. To this end, other supports have been investigated which do not have such drawbacks. Metal supports have been shown to be suitable substitutes, given their larger adhesion energy compared to polymers, making TMDs less prone to tearing. Lin et al. outlined a method by which TRT was used to transfer CVD-grown $MoS_2$ from a $SiO_2$/Si substrate to a target[44], which is shown schematically in **Figure 7**. A Cu thin film (~60 nm thick) was first coated on the $MoS_2$, to provide both a robust mechanical support and to separate the $MoS_2$ and the glue from the TRT. The TRT was placed on top and then peeled to remove the Cu/$MoS_2$ assembly from the substrate. The $MoS_2$ film was then transferred to a target substrate, and the TRT was released by heating to 120 °C. The Cu film was removed by etching.

Although the above method solved the polymer residue issue, it still led to cracks and holes in the transferred film. This was due predominantly to the mechanical strain incurred from peeling with TRT. Another method, utilizing a Cu support layer but without TRT, was developed by Lai et al. to avoid this.[105] In this method, they relied on water intercalation to delaminate the $MoS_2$ from its growth substrate. Use was made of a PDMS/PMMA layer on top of the Cu film on $MoS_2$ to help remove it after immersion in water. The buoyancy force supplied by the water was key in preventing damage to the thin PDMS/PMMA/Cu/MoS$_2$ assembly during peeling, as was the rigid Cu film support.

In general metal supports are more robust, but they share a similar drawback with polymer supports in that they require removal via chemical etching in the last transfer step, which can damage the TMD films. In addition, electron beam evaporation, although a softer metal deposition method than sputtering, can also damage the film. The process is also relatively expensive due to the used metal, restricting its use in industrial applications.

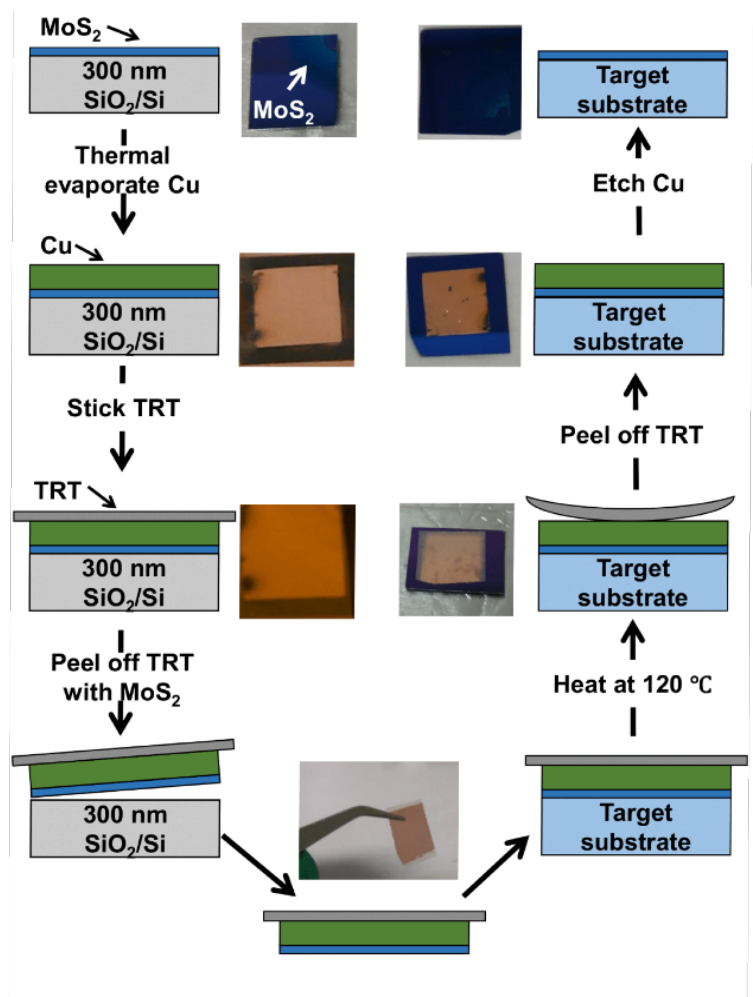

**Figure 7.** Illustration of the Cu-assisted transfer process. MoS$_2$ on SiO$_2$/Si is coated with a thin layer of Cu, and then TRT is placed on top. The TRT/Cu/MoS$_2$ assembly is peeled off the growth substrate, and placed on the target substrate. TRT is removed via heating, and the Cu is removed via etching, leaving the MoS$_2$ on the target substrate. Reproduced with permission.[44] Copyright 2015, Springer Nature.

## 2.2 Transfer without a supporting layer

Supporting layers, either polymer or otherwise, were introduced primarily to avoid damage to the 2DLM from the harsh etching solution used to detach it from the substrate, and to provide structural support in moving the film to the target substrate. Furthermore, in order to fully take advantage of the myriad exotic functionalities of 2D TMDs, it is necessary to be able to transfer them from their growth substrates to a variety of target substrates, in order to realize specific functionalities (e.g. flexible supports). Therefore, generic strategies that are not limited to specific substrates depending on the support are necessary moving forward. In this context, Xia et al., adapting a method used to transfer CVD-grown graphene[106], employed a direct transfer method to transfer $MoSe_2$ flakes to a TEM grid.[107] The procedure is outlined in **Figure 8**(a). A $MoSe_2$ film was grown using CVD on $SiO_2$/Si and mica substrates. These were attached to a TEM grid and then immersed in a 1% HF solution to etch the substrate, leaving the $MoSe_2$ film on the TEM grid. The fact that this technique removed the need of a support resulted in a faster and more convenient transfer, and did not require any post transfer chemical treatment. Nevertheless, despite the advantages of the aforementioned method, an etchant is still required to detach the 2DLM from the growth substrate, thereby limiting its industrial application as the growth substrate cannot be reused. To improve this, an all-water based transfer procedure for TMDs was developed, which did not use any harsh chemicals in any of the steps.[68,108,109] Kim et al. made use of such a method using centimeter-scale $MoS_2$ on $SiO_2$/Si as a representative case.[108] The steps are outlined in Figure 8(b). Instead of a chemical etchant, the $MoS_2$/$SiO_2$/Si assembly was immersed in water, which resulted in immediate delamination due to the difference between the hydrophobic $MoS_2$ and hydrophilic $SiO_2$ layer. The $MoS_2$ layer was subsequently scooped out of the water bath onto an arbitrary substrate. The growth substrate, having not been etched, could be recycled for another growth phase.

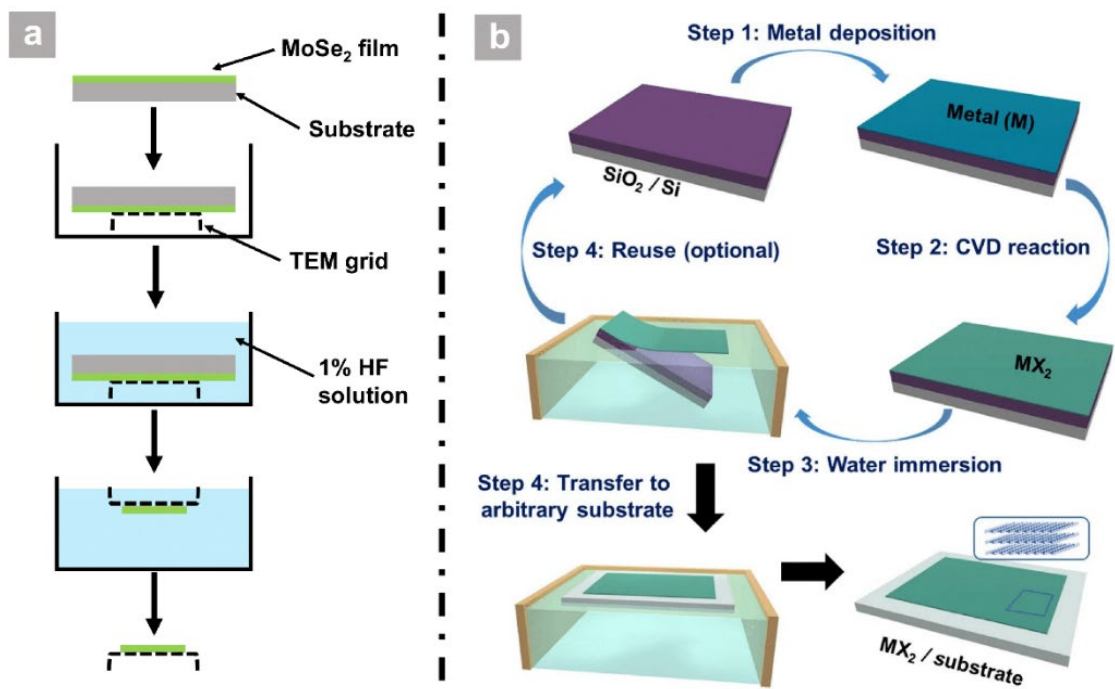

**Figure 8.** (a) An unsupported, direct transfer method using an aqueous solution. The MoSe$_2$ film on its growth substrate was placed in a container, with the MoSe$_2$ side contacting a TEM grid. The container was filled with 1% HF solution, resulting in delamination of the MoSe$_2$ film from its substrate. The MoSe$_2$, which remains on the TEM grid, floats to the surface of the solution, and can then be fished out and dried. Reproduced with permission.[107] Copyright 2014, Royal Society of Chemistry. (b) A schematic illustration of an all water-assisted transfer. The CVD-grown TMD can be immersed in water, resulting in immediate delamination from the growth substrate, and then fished out onto a target substrate. The growth substrate can be reused, as no chemicals are used to etch the surface. Reproduced with permission.[108] Copyright 2019, Springer Nature.

## 2.3 Discussion

The transfer methods of large-scale CVD-grown 2D TMDs have undergone significant development since their adoption from those used for graphene. The use of polymer supports became predominant owing to their flexibility and mechanical stability. However, from the above considerations it is clear that there is significant variation between polymers, and that the choice of polymer will invariably come with a need to adapt the transfer procedure to suit the polymer. Alternative (e.g. metal) supports exist, providing some advantages over polymers, and there are now methodologies which forgo the use of any support. There now exists a landscape of transfer methods for researchers to choose from, and this can be somewhat bewildering at first glance. It is the purpose of this review to give some clarity in this regard.

It is clear that the PMMA-assisted method, although the most extensively used, suffers from significant drawbacks. These include the use of harsh chemicals (such as KOH or HF) to etch the growth substrate, as well as the dissolution of the polymer after transfer, using hot acetone. These processes degrade the quality of the transferred TMD films. For example, the carrier mobility of CVD-grown monolayer $MoS_2$ can be reduced from ~8 $cm^2$ $V^{-1}$ $s^{-1}$ for the as-grown film[110], to 0.8 $cm^2$ $V^{-1}$ $s^{-1}$ for the PMMA-transferred one.[46] PDMS represents an improvement, in that it does not require removal via chemicals. However, delamination via mechanical peeling can result in an imperfect lift off, particularly given the poor adhesive properties of PDMS. This problem is often compounded by the strong film-substrate interactions that are introduced in the high-temperature growth process in CVD.[104] Water intercalation can assist in the delamination process, however it requires a hydrophilic layer under the as-grown TMD films (in contrast to the hydrophobic TMD), which is not common in the CVD growth-process. Substrates such as sapphire or $SiO_2$/Si (with the $SiO_2$ layer thickness greater than 300 nm[111]) have the necessary hydrophilic qualities. In the surface-energy assisted transfer described in ref[92], PS was used as a support. It was argued that, due to the greater hydrophobicity of PS, it can adhere more to the TMD layer than PMMA. Furthermore, the use of toluene to dissolve the PS layer resulted in a cleaner surface. Importantly, as mentioned above, PS is dissolved in THF to a greater degree than PMMA is in acetone.[94] This represents a potential solution to the problem of polymer residue, if the brittleness of PS can be addressed.

The addition of water-soluble polymers such as PVA or PVP represent an important modification to CVD transfer methods using supports, as they do not require any chemical solvents in the final transfer step. The entire process can be carried out using only water (see ref[104]), making this an environmentally friendly method. Water-based delamination was also central to the development of a transfer process that did not use any support (see ref[108]). The lack of support, however, can result in the wrinkling of the TMD film on the water surface. This process is not unlike how plastic kitchen film can wrinkle and fold without any supporting structure. It could be expected that the larger bending modulus of TMDs (and their associated resistance to crumpling) compared to graphene would be advantageous in transfers that use such support-free methods. Nevertheless, the water-assisted method requires a difference in surface energies between film and substrate, which again limits its use to specific substrates.

From one extreme of having no support, a method using a more robust support was developed. This was done to avoid the problem of the polymer film being very soft (low Young's modulus) compared to TMDs. For example, the Young's modulus of PMMA is around 8 MPa[34], much lower than that of $MoS_2$ (270 GPa)[25]. As a result, the polymer can fold in a manner outlined in **Figure 9**. Once transferred to the target substrate, these folds remain and there is reduced contact between the TMD and the surface. By comparison, Cu has a Young's modulus of 100 GPa[44], much closer to that of $MoS_2$ (and other TMDs), making for a much more robust support. The disadvantage of using metal supports, however, is that it still requires an etchant to remove the metal, damaging the TMD film. Furthermore, the use of e-beam evaporation to produce the thin metal support is relatively expensive, prohibiting such a method from being used in industrial applications. One way around this would be to use the metal foil growth substrate as the support as well. The problem is that the foils are quite thick (~100 μm[65]), making them less flexible than polymer films and resulting in poor contact with the 2DLM if they are bent. Reducing the thickness may address this limitation.

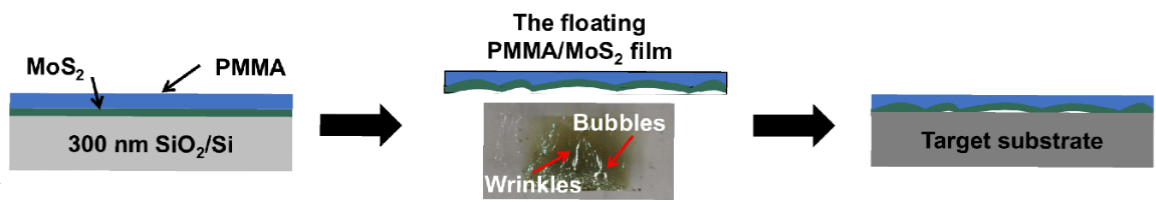

**Figure 9.** Schematic illustration of the wrinkle formation process in standard PMMA transfers. The photograph shows a PMMA/MoS$_2$ film floating on a KOH solution during transfer with TRT. Bubbles and wrinkles can both be seen. Reproduced with permission.[44] Copyright 2015, Springer Nature.

In the light of the various technological applications of TMDs and their stacked combinations, thought should also be given to the applicability of these transfer methods, not only with respect to scalability, but also in terms of target substrates. For instance, the roll-to-roll (R2R) method, described in Section 2.1.4, is scalable and suitable for flexible targets such as PET or EVA, but it is not applicable to inflexible SiO$_2$/Si substrates and therefore of little use in the semiconductor industry.[112] By comparison, water-soluble supports such as PVA or PVP can be used to transfer TMD layers to various substrates, including Cu foil, SiO$_2$/Si and quartz.[83]

To summarize, it is apparent that despite the significant developments in transfer methods for CVD-grown TMDs, notable challenges remain. Hence, more research is required on improving these methods, with an emphasis on industrial applicability. The criteria that must be met are as follows: (i) Reduced contamination (e.g. polymer residue) and TMD film degradation (e.g. wrinkles and cracks), (ii) cost-effective methods that allow for scalability, and (iii) wide applicability in terms of target substrate (particularly Si wafers for CMOS integration).

## 3. The impact of transfer techniques on film quality

Using the different kinds of transfer methods mentioned above, various TMDs can be successfully transferred onto a wide range of different substrates. Given the potential for using 2D TMDs for novel technological applications, it is vital that the transferred films are of a high quality and fidelity, in order to preserve the material properties. To this end, the structural, chemical and electronic properties of transferred TMD films have been investigated.[29,44,47,76,92,99,113–115] This is important as many transfer methodologies suffer from process related weaknesses, such as trapped bubbles, polymer residues, cracks or wrinkles. These features can degrade device performance. For instance, inhomogeneous or uncontrollable strain is detrimental to photoluminescence (PL) and optical applications[48,116,117], and cracks, wrinkles or polymer residue can strongly affect device resistivity and electron mobility[46,57,118]. It should be noted, however, that the various effects will be of differing importance depending on the application. For example, polymer residue does not have a large influence over the photoluminescence signal of 2D TMDs.[119] Thus, for optical applications such issues can usually be ignored. In the ideal case, a perfect transfer entails the functional continuity of the 2D film before and after transfer, with the only difference being the substrate. In practice this does not occur and, depending on the method used, modifications to the film results. In this section, the drawbacks associated with transferring CVD-grown TMD films will be outlined and described in detail. Furthermore, characterization techniques that can be used to quantify these drawbacks and methods to improve the film quality post-transfer will be given.

### 3.1 Issues encountered in transferring 2D semiconductors

*3.1.1 Wrinkles and cracks*

Wrinkles in 2D materials are to a certain extent unavoidable, as predicted by the Mermin-Wagner theorem.[120] Ultimately, long-wavelength fluctuations destroy the long range order of 2D crystals. Once the size in one dimension exceeds a critical value (of the order of nanometers for van der Waals materials), the material will wrinkle due to thermal fluctuations.[121] Substrates can strongly suppress this effect, meaning that these natural wrinkles of 2DLMs can be mitigated by coupling them to supports. In addition to this 'built-in' wrinkling, other wrinkled structures can form randomly during the CVD growth and transfer processes, in a manner which is unavoidable.[56] Cracking can also occur when applied stresses, such as those that occur during transfer, break the chemical bonds of the material. As mentioned in Section 2, 2D TMDs are more resistant to cracking than graphene due to their trilayer atomic structure. They have a higher fracture toughness[122], and during the CVD growth process they do not suffer from the stresses that occur due to having a negative TEC, as graphene does. Indeed, most materials, including TMDs and common CVD-growth substrates, have a positive TEC. Nevertheless, cracks and wrinkles still occur, either from the growth or transfer procedures. Although cracks in transferred films are undesirable, particularly for applications, this is not so obvious for the case of wrinkles. Indeed, wrinkle engineering can be used to tune the electronic properties of 2D materials, such as planar mobility.[123] For the case of $WS_2$, wrinkles greatly enhanced the intensity of the photoluminescence signal, via the tuning of the bandgap[124], which would find application in efficient photodetectors. On the other hand, both wrinkles and cracks can cause carrier scattering via flexural phonons[57], resulting in a lower mobility, as well as short circuiting, which damages device integration.[118] In such situations it is important to understand the origins of these structural modifications, and how to reduce them. Here we limit the discussion to the transfer process.

Wrinkles and cracks can come from various steps during transfer. For example, during the etching process to remove the growth substrate, some of the wrinkles formed from the surface topology of the growth substrate can be removed via the large surface energy of the etchant.[56] At the same time, the etchant may induce the formation of new wrinkles via capillary forces. A soft support layer (i.e. one with a low Young's modulus compared to the 2DLM) cannot prevent deformation of the film after it is detached from its growth substrate, for thicknesses in the order of hundreds of microns. These wrinkles remain when transferred onto the target substrate, reducing the direct contact area of the film with the surface. The evaporation of the solvent used to remove the polymer can also induce excess wrinkling.[125] In addition, methods that rely on mechanical peeling to remove the polymer layer induce a lateral strain in the 2D film when the polymer/TMD assembly is pressed onto the target substrate. When the polymer is peeled off, this can damage the TMD layer.[126]

Gurarslan et al. compared the surface-energy-assisted method using a PS support, to that of a conventional PMMA-assisted one, for transferring CVD-grown $MoS_2$ (see Section 2.1.3).[92] The PMMA/$MoS_2$ assembly on sapphire was immersed in hot (90 °C) NaOH solution for 4 hours during the transfer process. Holes and cracks were observed in the transferred $MoS_2$ film using the conventional PMMA support, as shown in **Figure 10**(a). By comparison, the surface-energy-assisted PS-support method produced no observable cracks or wrinkles. The reasons for this are two-fold. Firstly, the use of hot chemical etchants produces bubbles that can be trapped between the film and support layer, inducing mechanical strain and causing folding or even cracking. The use of water at room-temperature can help to alleviate these effects. Secondly, PS provides a more robust support than PMMA due to its larger Young's modulus, hence limiting wrinkle formation.

Bubbling in a solution[65] and/or by the capillary force induced when transferring the film out of the solution or water bath[108] (the wedging transfer method) can also result in wrinkles and cracks. For instance, Figures 10(b) and (c) show the optical images of transferred monolayer $WS_2$ from Au foil onto $SiO_2$/Si wafer by a PMMA-based wet-etching method, and the bubbling transfer method as described in Section 2.1.1, respectively.[65] It can be seen that the coverage of $WS_2$ via the wet-etching transfer is higher than that from the bubble transfer (see also Figure S11 of the Supplementary in ref[65]). This indicates the extra mechanical stress on the TMD film by the intercalated bubbles. Figure 10(d) shows an atomic force microscopy (AFM) image of the same transferred WS$_2$ film on the SiO$_2$/Si substrate via the bubble method (as shown in Figure 3(a)). Wrinkles can be clearly observed. Similarly, Figure 10(e) shows an AFM image of monolayer MoS$_2$ transferred from Au foil to SiO$_2$/Si via wet-etching of the supporting PMMA layer (in addition to an Au etchant, KI/I$_2$).[127] Cracks as well as wrinkles were observed. These findings indicate that wrinkles can occur at many stages of the transfer process, and their removal can be challenging.

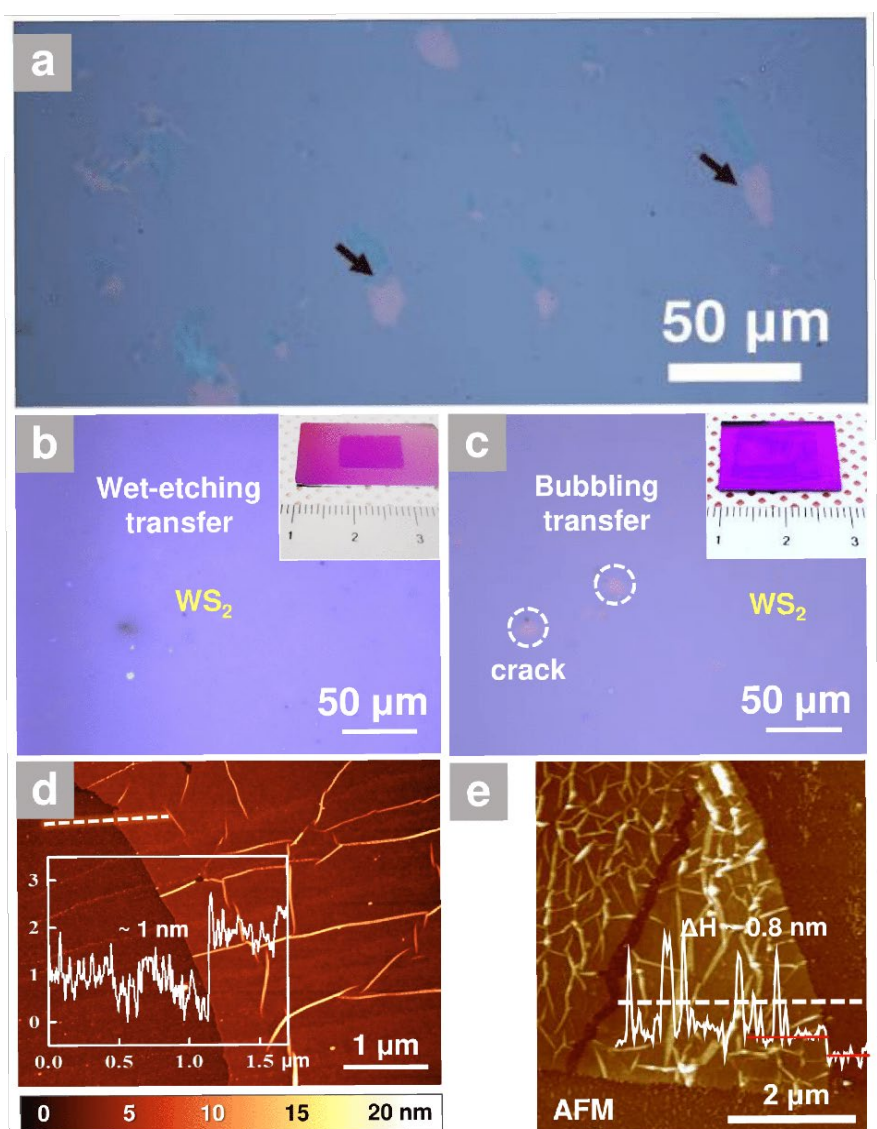

**Figure 10.** (a) Optical image of transferred MoS$_2$ on SiO$_2$/Si. The arrows indicate gaps in the film. (b-c) Optical images of transferred WS$_2$ on SiO$_2$/Si via wet-etching, PMMA-assisted and bubbling transfer, respectively. Cracks and gaps are more clearly observed in the bubble method. (d) AFM image of the WS$_2$ film transferred using the bubble method. Wrinkles are clearly observed. (e) AFM image of MoS$_2$ transferred via the PMMA-assisted wet etching method, showing again significant wrinkling. (a) Reproduced with permission.[92] Copyright 2014, American Chemical Society. (b-d) Reproduced with permission.[65] Copyright 2015, American Chemical Society. (e) Reproduced with permission.[127] Copyright 2015, John Wiley & Sons, Inc.

*3.1.2 Bubbles at the interface between TMD and substrate*

The transfer process consists of two main steps: (1) the removal of the 2D film from the growth substrate, and (2) the placing of the film onto the target substrate. The latter step is prone to trap contaminants at the interface formed between the TMD and the target substrate. However, due to the high diffusivity of contaminants on 2D van der Waals crystals (also termed the "self-cleansing" mechanism[128]), these contaminants tend to aggregate into bubbles or blisters. This occurs because of the difference in adhesion energy between the TMD and the target substrate, and the TMD and contaminants. If the former is larger, then it is energetically favorable for the two materials to have the largest possible interface. This has the effect of pushing the contaminants away, leading to the aggregation. This is shown clearly in **Figure 11**, in which AFM images were taken of graphene transferred onto various 2D crystals. On substrates with a good adhesion to graphene, such as hexagonal boron nitride (h-BN), MoS$_2$ and WS$_2$ (Figure 11(a-c)), the contaminants are observed to aggregate into bubbles. On the other hand, on substrates with a poor adhesion to graphene, such as mica, bismuth strontium calcium copper oxide (BSCCO), and vanadium oxide (V$_2$O$_5$) (Figure 11d-f), the contamination is observed to spread uniformly over the interface.

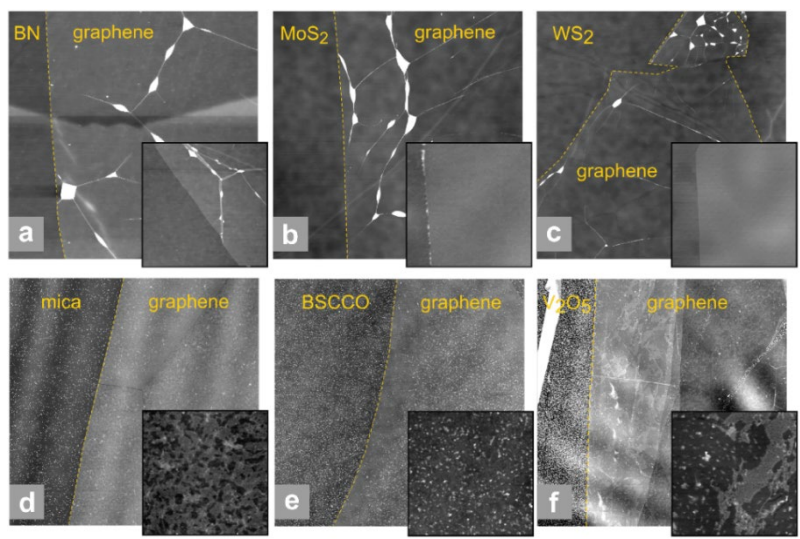

**Figure 11.** AFM images of graphene on various substrates. In images (a-c), graphene is placed on top of van der Waals surfaces (h-BN, MoS$_2$ and WS$_2$). It can be seen that, due to the self-cleansing mechanism of 2D van der Waals materials, large areas of graphene/substrate interface become flat and contaminant free, with a surface roughness on the order of 0.1 nm. Contaminants are seen to aggregate into bubbles or blisters. In images (e-g), graphene is placed on hydrophilic oxide surfaces. It is observed that no large bubbles are present, with a surface roughness of a few nm. Reproduced with permission.[128] Copyright 2014, American Chemical Society.

Trapped water or residue from chemical etchants are mainly found in wet transfer methods[68,76,129,130], which can remain on a surface and thereafter become trapped between the transferred TMD film and that surface (for instance the supporting layer during transfer, or the target substrate). However, contaminants can also be introduced during all-dry transfers, including trapped air pockets.[68,131] It has been previously observed that bubbles formed during PMMA-assisted transfers contained amorphous hydrocarbons, as would be expected of PMMA contamination.[132] Little is known about contaminants introduced in other transfer methods. Hong et al. transferred monolayer MoS$_2$ flakes from soda-lime glass onto HOPG (highly oriented pyrolytic graphite) by a water-assisted method (without support), and an all-dry TRT-assisted method.[68] **Figure 12**(a-b) shows scanning electron microscopy (SEM) images of the wet-transferred sample, where clear bubbles are observed. These are attributed by the authors to trapped water at the interface of $MoS_2$ and HOPG. The scanning tunneling microscope (STM) image in Figure 12(c) indicates some surface inhomogeneity. The reduction in the number and size of the bubbles from the dry-transferred sample, shown in the SEM images in Figure 12(d, e), confirms the origin of the bubbles in the wet-transferred case. Nevertheless, even in the all-dry case bubbles can be seen, which were attributed primarily to trapped air.

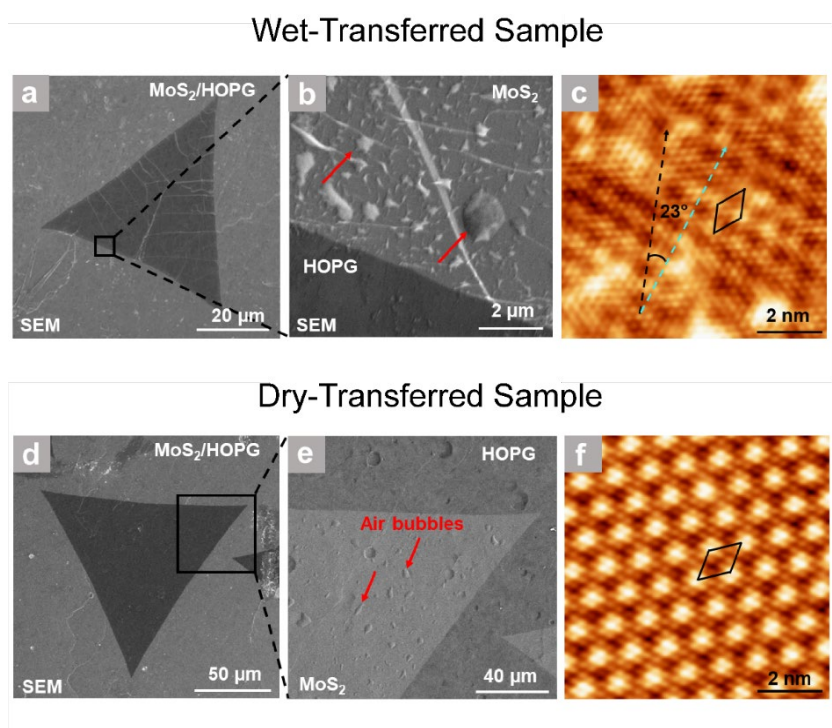

**Figure 12.** A comparison between an all-water assisted wet method (a-c) and TRT-assisted all-dry method (d-f) to transfer $MoS_2$ from soda-lime glass to HOPG. (a,b) SEM images of an $MoS_2$ flake on HOPG after the water-assisted transfer process. Numerous bubbles and wrinkles can be observed. (c) STM image of the same flake as in (a), showing the rhombic unit cell of the Moiré lattice, with randomly distributed bright spots. (d,e) SEM images of an $MoS_2$ flake on HOPG transferred using the all-dry TRT-assisted method. A few air bubbles can be observed. (f) STM image showing the same Moiré lattice, without the inhomogeneous distribution of bright spots. Reproduced with permission.[68] Copyright 2018, John Wiley & Sons, Inc.

The issue of bubbles/blisters is particularly pertinent to TMD-heterostructures. A common fabrication route for such heterostructures involves targeted pick-up and release. These stacking methods are known to introduce contaminants and prevent the formation of pristine interfaces, which can affect interlayer interactions (such as charge and energy transfer).[133,134] For instance, Yang et al. observed bubbles when fabricating graphene/TMD heterostructures using a van der Waals pick-up method.[134] The method is shown schematically in **Figure 13**(a). An exfoliated flake of h-BN was picked up by a PDMS-based support, and was then brought into contact with a CVD-grown $WSe_2$ layer on $SiO_2$/Si. In the final step (not shown) the polymer/h-BN/TMD assembly was brought into contact with graphene, completing the heterostructure. It was found that bubbles were mainly introduced in this step. From the AFM image in Figure 13(b) it can be seen that the bubbles form randomly, and the PL mapping in Figure 13(c) of the same area as in Figure 13(b) indicates red dots where the PL signal is not quenched. In this respect, PL serves as a helpful tool for characterizing bubbles, as the PL signal is quenched when graphene is brought into intimate contact with the TMD.

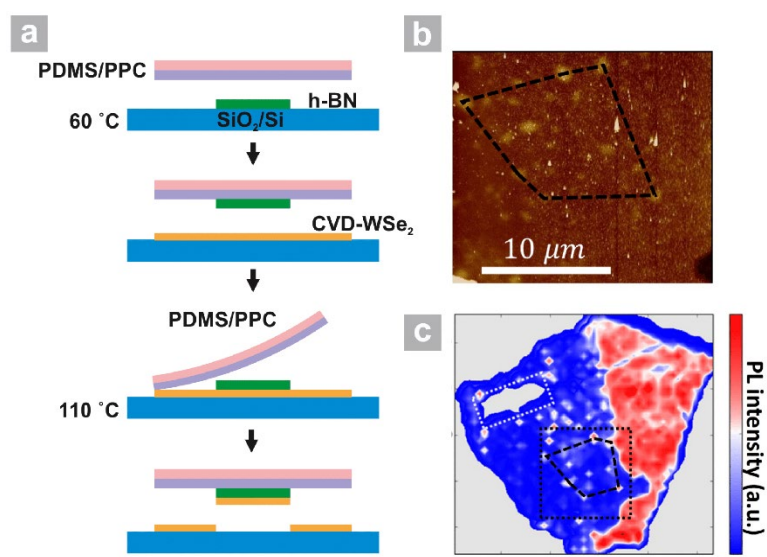

**Figure 13.** (a) Procedure for transferring CVD-grown $WSe_2$ using an h-BN flake, with a PDMS-based all-polymer support. Firstly, a PDMS/PPC (polypropylene carbonate) stamp is lowered towards an h-BN flake (with the sample kept at 60 °C). The h-BN is detached from the substrate, and is then brought into contact with a CVD-$WSe_2$ film. The PDMS/PPC/h-BN/$WSe_2$

assembly is then peeled from the substrate, and is ready to align onto a pre-exfoliated graphene flake. (b) AFM topography image of the resulting h-BN/TMD/graphene stack, showing bright protrusions corresponding to the bubbles, imaged in (c) using PL mapping. Reproduced with permission.[134] Copyright 2017, American Physical Society.

As mentioned before, the existence of bubbles precludes a pristine interface, which is instrumental for emerging phenomena in van der Waals heterostructures, such as proximity effects and interlayer excitons. The self-cleansing mechanism of 2DLMs can lead to bubble-free regions over which devices can be constructed, however at larger scales this is not possible due to the layer size. Thus, solutions must be found to reduce interface bubbles. In theory, the presence of bubbles can be removed by controlling the angle at which the 2DLM is brought into contact with the target substrate, as well as the merging time. With a slower merging time at an angle other than normal incidence, bubbles have a greater chance of escape. This process is analogous to how a plastic screen protector adheres to a phone screen. Nevertheless, it is difficult to entirely remove such defects. It is therefore a matter of future research to address these issues.

*3.1.3 Residues from support layer*

Almost all transfer methods require a support to successfully transfer the TMD film as a continuous piece, maintaining uniformity. Section 2 outlined two types of supporting layers that have been used, namely polymer and metal. Polymer supports are used more frequently, in particular PMMA (and to a lesser extent PDMS). These supports tend to leave residues on the surface of the 2DLM after they have been removed. For instance, due to the strong dipole interactions between PMMA and graphene, a thin layer of PMMA remains on the surface after transfer and removal of the polymer.[97,135–138] **Figure 14** shows CVD-grown $MoS_2$ flakes transferred from a SiO$_2$/Si substrate to a target substrate using a PMMA-assisted transfer method. Figure 14(a) and (b) shows optical images of the as-grown and transferred MoS$_2$ flakes, respectively. Large residues can be seen in the latter image, on both the substrate and TMD, and these are confirmed by the AFM image in Figure 14(c). Such residue is known to degrade the intrinsic properties of 2DLMs. For example, it can decrease the mobility of graphene by more than 50% due to carrier scattering[139,140], and decreases its thermal conductivity by 70% because of phonon scattering[141]. Furthermore, such residue has been observed to cause weak p-type doping in transferred graphene, which can shift the threshold voltage for back-gated graphene FETs.[142] Research on the effects of PMMA residues on transferred TMDs is comparatively rarer.

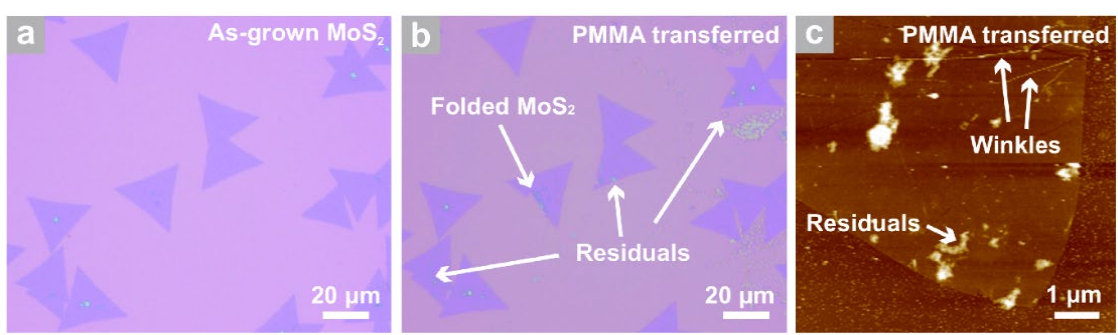

**Figure 14.** (a, b) Optical microscopy images of monolayer MoS$_2$ before and after transfer by PMMA-assisted transfer method on the SiO$_2$/Si. PMMA residuals are notable on the surface of transferred MoS$_2$ flakes. (c) AFM image of transferred MoS$_2$, the residues are clear on top of the MoS$_2$ flakes. Reproduced with permission.[44] Copyright 2015, Springer Nature.

A number of different methods have been employed to reduce or remove the residue. Annealing in different atmospheric conditions or in vacuum is the most common way[143-145], although the process is not completely effective[97]. Moreover, high temperature annealing may induce defects in TMDs, such as metal or chalcogen vacancies.[146,147] Annealing graphene samples in oxidative atmospheres to remove PMMA residues has been reported[148], but extending this method to TMDs is likely not a good idea because it may lead to oxidation. Laser cleaning and electrostatic-force cleaning of PMMA residues on graphene has been reported to have some success, although such methods have not yet been reported for TMD transfer.[140,149] Plasma cleaning has also been used[150], although care must be taken. For instance, $O_2$ plasma can lead to significant doping of TMDs.[151]

In addition to these global treatment methods, more local methods have been tried to remove PMMA residue. These include the use of contact mode AFM, where the residue is swept away by the tip to clean a small area of the film surface. For example, Liang et al.[152] fabricated $MoS_2$ and $WSe_2$ FETs by electron beam (e-beam) lithography (using PMMA as e-beam resist) and investigated the impact of post-lithography PMMA residue on the electrical characteristics of the two FETs. Using an AFM tip, they managed to lower the height topography of the surface, and found that the charge carrier density and source-drain current increased by $4.5 \times 10^{12}$ cm$^{-2}$ and 247%, respectively. It should be noted that such methods are clearly not scalable, and are thus not suitable long-term solutions to the residue problem.

Polymer residues are also found in PDMS-assisted transfer methods. PDMS, as mentioned in Section 2.1.2, contains many uncrosslinked oligomers which can remain on the surface after the polymer layer is detached after transfer, causing contamination.[29,86,115] On the one hand, such contamination reduces the surface cleanliness, affecting the properties of TMD heterostructures.[29,76] On the other hand, the transferred PDMS oligomers could be used as a protective layer for selected areas to survive from chemical etching.[153] Moreover, a patterned PDMS stamp can selectively pattern a target substrate with transferred PDMS oligomers to fabricate transistor devices.[154] To date, the influence of PDMS residues on the physicochemical properties of transferred TMDs has not been extensively researched. A possible reason would be that PDMS residues do not affect the overall PL quantum yield, and thus has been overlooked.[115] Various methods have been employed to remove PDMS residue, some of which overlap with those for PMMA. Annealing at 400 °C under ultra-high vacuum (UHV) conditions for 1 hour has been reported to completely remove PDMS residues on MoS$_2$[115], however such high temperatures can introduce defects. Dissolving the PDMS residues in organic solvents, such as acetone or hexane, has been shown to be effective.[155] Usually, the PDMS swells when it is in organic solvents and the amount of extracted PDMS oligomers increases as the swelling ratio increases.[156] However, the solvent molecules might be adsorbed on the transferred TMD surface and perhaps lead to chemical doping. Preemptive treatment methods have also been investigated, for instance by pre-cleaning the PDMS by ultraviolet/ozone (UV/O$_3$).[115,157] Jain et al. employed such a step for MoS$_2$ flakes exfoliated on PDMS and transferred onto h-BN on a SiO$_2$/Si substrate.[115] They found that the amount of PDMS residue was significantly reduced, as shown in **Figure 15**.

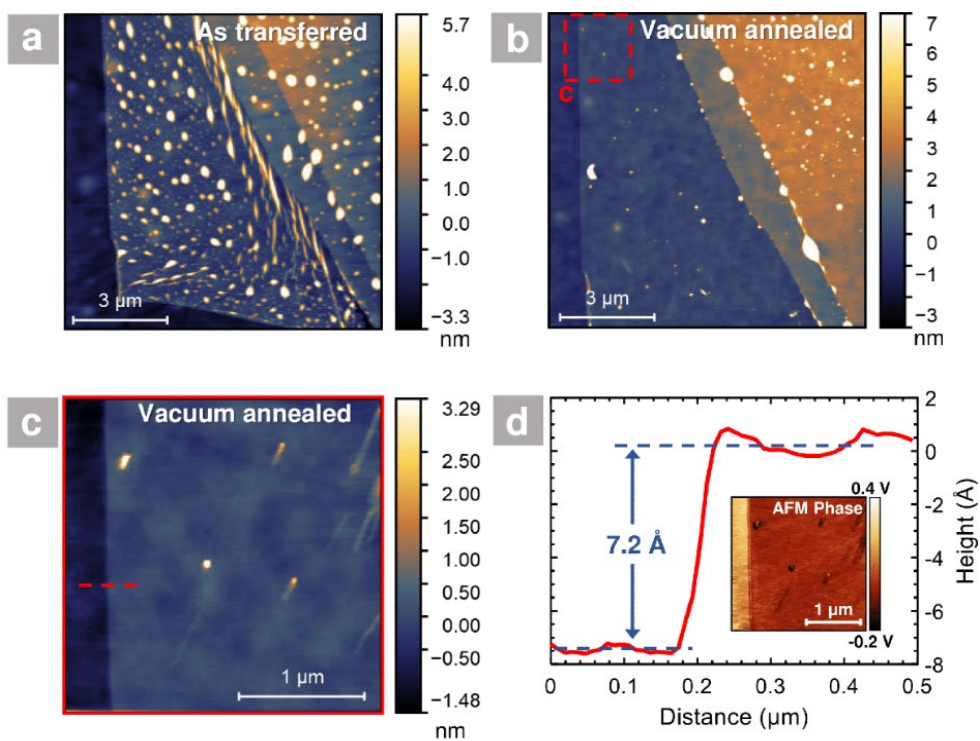

**Figure 15.** (a) AFM image of transferred MoS$_2$ on h-BN, displaying a considerable number of bubbles and wrinkles. (b) AFM image of the same area in (a), showing that the bubbles were efficiently removed by vacuum annealing at 200 °C for 3 hours. (c) AFM image of the red-outlined area in (b). (d) Height profile along the red-dashed line in (c), exhibiting a clear

monolayer MoS$_2$ step of 7.2 Å. Inset: AFM phase map recorded together with the topography in (b) revealing a clear phase contrast between MoS$_2$ and h-BN. Reproduced with permission.[115] Copyright 2018, IOP Publishing.

**3.2 Characterization techniques for determining transferred film quality**

Given the numerous issues relating to the quality of transferred CVD-grown TMDs, and the need to resolve them for future applications, a robust set of characterization tools are required to benchmark the transferred films, at various length scales ranging from the atomic to the macroscopic. Here an overview of the main techniques is presented, as well as less common techniques that may also find utility in such inquiries.

*3.2.1 Raman spectroscopy*

One of the most common and readily accessible methods for obtaining structural and chemical information of materials is that of Raman spectroscopy. As a relatively cheap and non-destructive technique, it has played an important role in the characterization of graphitic materials. As a result, it has been readily adopted to study the quality of transferred TMD films. A typical Raman spectrum for TMD films can yield information on the number of layers[19,44,92,158,159], indicate the charge doping[44,68,130], strain[160] and defect density[161,162]. The Raman spectra of 2D TMDs draw some comparisons to graphene, with both similarities and notable differences. Semiconducting TMDs, like MoS$_2$ and WSe$_2$, often appear in the 2H phase. The 1T or 1T' phase is a meta-stable phase resulting in (semi)metallic behavior and only stable for a selected number of TMDs. The 2H phase belongs to the P6$_3$/*mmc* nonsymmorphic space group (D$^4_{6h}$), with an inversion symmetry between the two adjacent monolayers (shown in **Figure 16**(a)). This symmetry is shared by Bernal stacked graphite.[162] However, in the monolayer limit 2H-TMDs lose their inversion symmetry, reducing the space group to the symmorphic $P\bar{6}m2$ ($D^1_{3h}$).[163] Graphene has only one first-order (doubly degenerate) Raman active mode belonging to the irreducible representation $E_{2g}$. This mode gives rise to the so-called G band (1580 cm$^{-1}$). In contrast, monolayer 2H-TMDs have three Raman active modes corresponding to the $A_1'$, $E'$ and $E''$ irreducible representations. These are shown in Figure 16(a). $A_1'$ corresponds to the chalcogen atoms vibrating in the out of plane direction, with the upper chalcogen atom moving in anti-phase with the lower one. The metal atom remains stationary. The $E'$ and $E''$ modes correspond to in-plane atomic vibrations. For the $E'$ modes, the metal atom moves in anti-phase with the two chalcogen atoms. It is somewhat comparable to the $E_{2g}$ mode in graphene. In a standard Raman backscattering configuration, the $E''$ mode is silent and thus a typical Raman spectrum of monolayer 2H-TMDs displays only 2 first-order Raman modes ($A_1'$ and $E'$)[164], although it should be noted that this spectrum is strongly dependent on the excitation wavelength of the light. A resonant excitation leads to a diverse number of second-order peaks due to strong electron-phonon coupling.[165]

In addition to the Raman active modes in pristine 2H-TMDs, there also exist defect-activated modes which can be useful for determining transferred film quality. Prominent defect-induced Raman bands were found in the spectrum of $MoS_2$ and $WS_2$ films at ~223 and 178 cm$^{-1}$, respectively.[166] They originate from phonons at the M-point of the longitudinal acoustic (LA) branch of the Brillouin zones of each material. For $MoS_2$, the intensity of this LA(M) peak was found to be proportional to the average distance between defects.[161] The underlying mechanism could involve a so-called double-resonance (DR) Raman process involving 1 phonon and a defect. This process involves the same phonon branch as another second-order Raman process, the 2LA(M) band in $MoS_2$ and $WS_2$, which involves the scattering by two phonons, in contrast to the one phonon plus a defect in the LA(M) band. The 2LA(M) and the defect-induced LA(M) bands are analogous to the transverse optical (TO) modes of graphene involving phonons at the K-point – 2TO(K) and TO(K). These two bands are also called the G' and D bands (or 2D and D bands, respectively), with the latter being associated with defects in graphene.

Raman spectroscopy can be used to determine the number of layers of transferred $MoS_2$ films. The $E^1_{2g}$ and $A_{1g}$ are sensitive to layer thickness, and in general the Raman shift of the former will decrease and that of the latter will increase with increasing layer number.[167] An example is shown in Figure 16(b), showing Raman spectra of the CVD-grown $MoS_2$ before and after transfer, of both monolayer and trilayer samples. The change in wavenumber ($\Delta k$) was found to be ~20 cm$^{-1}$ and 23.2 cm$^{-1}$, respectively.[92] Strain is another factor that influences the Raman spectra of TMD films. It was found that the change in wavenumber per percent of strain for uniaxially strained monolayer $MoS_2$ was -2.1 cm$^{-1}$ for the $E^1_{2g}$ (E') mode and -0.4 cm$^{-1}$ for the $A_{1g}$ ($A_1$') mode, in good agreement with theoretical predictions.[168] Such a redshift was observed in the $E^1_{2g}$ peak for CVD-grown $MoS_2$ after transfer with PMMA (~2 cm$^{-1}$)[44], as shown in Figure 16(c). Little change was observed for the $A_{1g}$ peak, in accordance with the low value reported in ref[169]. The authors ascribed the change in the $E^1_{2g}$ mode to wrinkle-induced strain after the transfer. By contrast, no change in the $E^1_{2g}$ mode is observed for the TRT-assisted transfer. However, a blueshift of the $A_{1g}$ mode was observed, which is generally associated with p-doping of the $MoS_2$ from charged impurities[168] introduced at the interface between the $MoS_2$ and the target substrate during transfer. Such behavior was not observed in the PMMA-assisted transfer due to the strain dominating the peak shifts.[170] Other than impurities, different substrates[47], substrate-borne moisture[171] and vacancy defects (such as sulfur)[68] can also lead to doping and thus change the $A_{1g}$ peak.

Raman spectroscopy is also a good way to study the interlayer coupling in van der Waals stacked bilayer TMDs, such as $MoS_2/WS_2$[29,76,129] and $MoS_2/WSe_2$[130]. For instance, it was found that in $MoS_2/WS_2$ heterostructures the E' and $A_1$' modes of each layer contributed to the Raman signal independently at the same frequencies as the individual monolayers (both before and after annealing), implying minimal interlaying coupling.[76] By comparison, MoS$_2$/WSe$_2$ heterostructures displayed a significant change in the Raman spectrum after thermal annealing.[130] The layer-sensitive mode A$^2_{1g}$ for WSe$_2$ at 309 cm$^{-1}$ was observed, and the E' and A$_1$' degenerate modes of WSe$_2$, as well as the A$_1$' of MoS$_2$, became blue shifted, with the E' mode of MoS$_2$ displaying a redshift. Lastly, Raman mapping has been extensively used to investigate the quality of transferred TMD samples, especially for checking the cleanliness and uniformity of the sample. Figure 16(d) shows an optical image of a ~100 μm sized monolayer WS$_2$ flake transferred onto a SiO$_2$/Si substrate by the bubbling transfer method (in NaOH solution), and below is the Raman map of the same sample. The map is relatively homogenous across the whole flake, indicating that the sample maintains good uniformity and continuity after transfer.

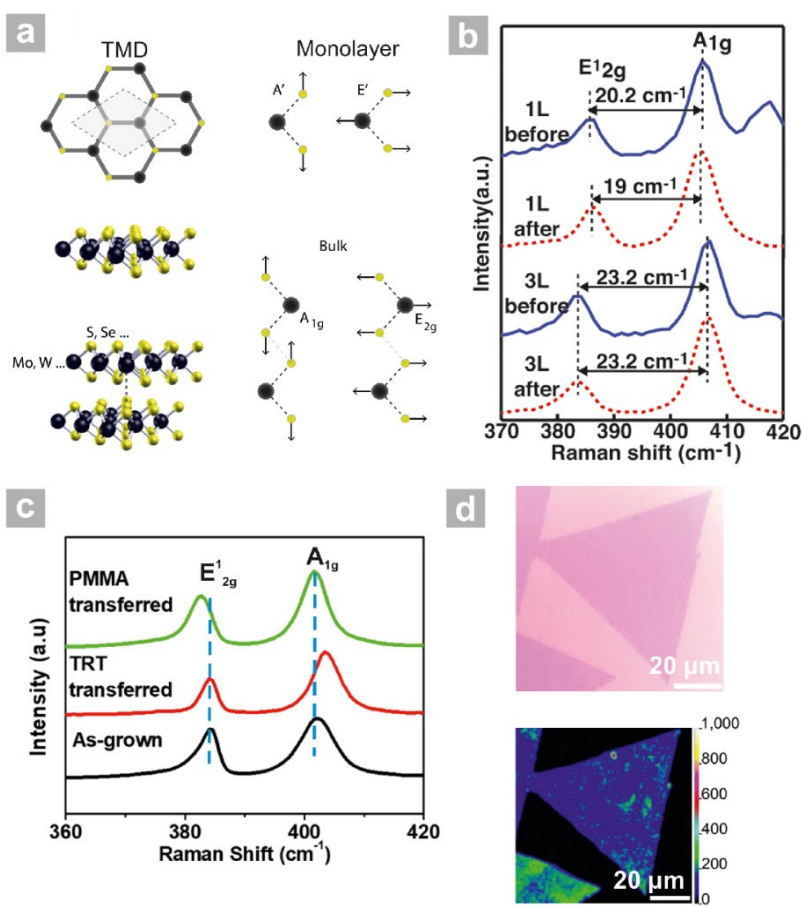

**Figure 16.** (a) Atomic structure of TMDs and Raman active modes of 1H or 2H TMDs. Reproduced with permission.[162] Copyright 2015, American Chemical Society. (b) Raman

spectra of the as-grown MoS$_2$ on sapphire before and after transfer onto SiO$_2$/Si by a PS-assisted transfer method. Reproduced with permission.[92] Copyright 2014, American Chemical Society. (c) Raman spectra of as-grown MoS$_2$ on SiO$_2$/Si, as well as transferred MoS$_2$ on SiO$_2$/Si using a TRT/Cu assisted and PMMA-assisted method. Reproduced with permission[44]. Copyright 2015, Springer Nature. (d) Optical microscopy image of transferred monolayer WS$_2$ on SiO$_2$/Si by bubbling transfer method. (e) Raman map of the intensity of A$_{1g}$ peak of the WS$_2$ flake in (d). Reproduced with permission.[172] Copyright 2015, Springer Nature.

*3.2.2 Photoluminescence spectroscopy*

A complementary optical characterization technique to Raman spectroscopy for TMDs is PL spectroscopy. As both techniques use a laser-source for excitation, they can be integrated into a single setup for sample characterization. It is well-known that MoS$_2$-type TMDs show a transition from an indirect to direct bandgap in the ML limit. This crossover is accompanied by a large increase in the PL signal as a result of direct excitonic transitions at the *K* (and *K'*) point of the Brillouin zone.[173] For MoS$_2$, this increase in PL intensity can be as large as 10$^4$ when compared to bulk.[174] Thus, PL is sensitive to the electronic structure in TMDs and thus monolayer thicknesses can be easily identified. Moreover, variations in the PL signal can provide information on film quality. These include layer number[48,174], charge-doping[65,175], strain[170,176-178] and defects[179-181].

The PL spectrum of semiconducting ML 2H-TMDs, such as MoS$_2$, features two main peaks. These are the so-called A and B excitons, and are the result of the spin-splitting of the valence (and to a lesser extent the conduction) band due to the strong spin-orbit coupling of the transition metal. This is shown schematically in **Figure 17**(a). The spin-splitting at the *K* (and *K'*) point in the valence band (VB) of Mo and W-based 2H-TMDs is about 0.2 eV and 0.4 eV, respectively.[182] In addition, the PL spectrum of TMDs can display additional many-body excitonic effects, such as trions, biexciton and trion-exciton complexes.[183,184] As with Raman spectroscopy, PL can provide information on defects in TMD films. McCreary et al. analyzed the room temperature PL of CVD-grown $MoS_2$, $MoSe_2$, $WS_2$ and $WSe_2$ monolayers and determined that PL variations arise from differences in the non-radiative recombination associated with defect densities.[179] The relative intensities of the A and B exciton emission peaks can be used to check sample quality; a low B/A ratio indicates low defect density (high sample quality), and vice versa.

The change in the PL peak position and strength can also indicate strain. Tensile strain for ML TMDs is introduced during the CVD growth process due to the difference in TEC between TMDs and the growth substrate. When transferred to a target substrate, this strain can be released, although the formation of wrinkles and trapped bubbles can also result in new sources of strain. Tensile strain can modulate the band structure of ML TMDs such as $MoS_2$, which affects the PL spectra.[176] Liu et al. transferred CVD-grown monolayer $MoS_2$ onto a PDMS substrate and applied uniaxial force to the sample.[177] It was found that the PL signal decreased with increasing strain, in an approximately linear fashion (see Figure 17(b)). Furthermore, the center of the peak was redshifted with increasing strain, due to a reduction of the optical bandgap.

Charge doping is another factor that strongly influences the PL of TMDs. ML TMDs can be doped in many ways, for instance through the chemi- or physisorption of electron acceptor or donor molecules, and also through structural defects (e.g. sulfur vacancies), substrate interactions, and polymer supports.[185-187] For example, Mouri et al. studied the PL properties of exfoliated monolayer $MoS_2$ via chemical doping.[185] The PL intensity was noticeably enhanced by p-type doping, and reduced with n-type doping (see Figure 17(c,d)). The former effect can be understood as a shift from trion recombination to exciton recombination, after the extraction of the residual electron. The latter can be understood as a suppression of the exciton contribution to the PL signal via electron injection. Localized n-doping due to charged structural defects can also be inferred from changes to the PL signal, as was reported by Peimyoo et al., in which a blueshift in the A exciton peak was observed in the PL signal of CVD-grown $WS_2$.[188] The substrate itself can also have an effect on the PL signal. Yu et al. investigated the substrate influence on PL of CVD-grown TMDs by transferring them onto various substrates.[189] They found that the main influence of the substrate is to dope the TMD, and to promote defect-assisted nonradiative exciton recombinations, while strain and dielectric screening contribute to a lesser extent. Suitable substrate choice could lower the doping effect, for instance mica for $WS_2$ and $MoS_2$, and h-BN or PS for $WSe_2$.[189]

Lastly, interlayer coupling in TMD heterostructures can be assessed via PL spectroscopy. For instance, Chiu et al. fabricated a $MoS_2/WSe_2$ heterostructure and observed a new PL peak at 1.59 eV after the sample was thermally annealed, which was attributed to the interlayer radiative recombination of spatially separated carriers, i.e. interlayer excitons.[130] Thermal annealing was also found to tune the interlayer coupling in $WS_2/MoS_2$ heterostructures, with a new PL peak emerging at 1.94 eV.[76]

PL spectroscopy provides complementary information to Raman spectroscopy on the quality of TMD films, but it also yields some advantages. Using Raman spectroscopy to characterize TMD defects at room temperature show only modest broadening of the FWHM of the E' and $A_1$' modes.[190] This is compounded by the fact that strain and doping also result in a shift and broadening of the characteristic peaks[176,191], and these changes are small and difficult to differentiate. Whilst PL spectroscopy also suffers from the ambiguity in assigning the origin of changes to characteristic peaks at room temperature, its strength lies in its marked temperature dependence. The PL signal can be enhanced via defect-confined carriers, with low temperature PL often showing clear defect-related PL peaks (known as $L$ and $L_H$).[192] These peaks are often difficult to detect at room temperature, since they show a sharp decrease in the intensity as the temperature increases, as shown in Figure 17(e). Thus, the stronger temperature sensitivity of PL over Raman confers an advantage of PL over that of Raman spectroscopy.[180]

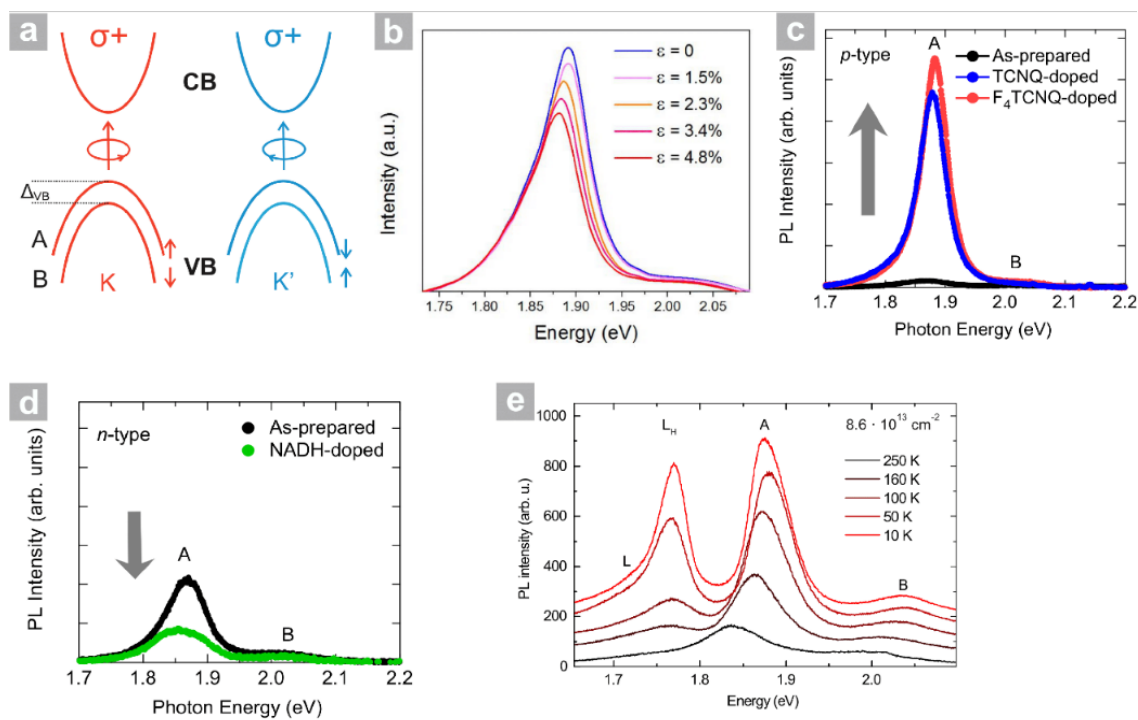

**Figure 17.** (a) Schematic illustration of the single particle electronic band structure, showing the spin splitting at the K and K' points due to the spin-orbit interaction. The valley-dependent optical selection rules are also shown, as well as the A- and B-exciton emission states. (b) Representative PL spectra from a CVD-grown ML $MoS_2$ flake under different tensile strain. From. (c) PL spectra of ML $MoS_2$ before and after p-type doping with TCNQ and $F_4TCNQ$ molecules, and (d) before and after being n-type doped with the molecule NADH. (e) Temperature dependent PL spectra at a $He^+$ defect dosage of $8.6 \times 10^{13}$ $cm^{-2}$ (using a helium ion microscope). The A and B excitonic emission peaks can be observed, as well as the defect related $L$- and $L_H$-peaks. (a) Reproduced with permission.[179] Copyright 2018, AIP Publishing LLC. (b) Reproduced with permission.[177] Copyright 2014, Springer Nature. (c, d) Reproduced with permission.[185] Copyright 2013, American Chemical Society. (e) Reproduced with permission.[192] Copyright 2018, IOP Publishing.

*3.2.3 Scanning probe microscopy*

Whilst Raman and PL spectroscopy provide information on the structural, chemical and electronic properties of TMDs, they do not yield any insight into surface topography. Features that are generally undesirable in transferred films, such as bubbles, wrinkles and polymer residues (see Section 3.1) also need to be characterized. To this end, scanning probe microscopy (SPM), a branch of microscopy that uses a physical probe to image the surface of a sample, is a versatile technique that can yield direct information on these undesirable surface features. Within the family of SPM, AFM and STM are the two most well-known and often used techniques. They can provide very high resolution, down to the atomic scale.

AFM is extensively used to study transferred TMD films because it is relatively easy to operate and has various imaging modes which provides distinct but complementary information on surface topography. Wrinkles, cracks, bubbles or polymer residues can be readily observed, and the height profile along the edge of the 2D flakes can give information on layer number[44,47,65,127,159,193], although this is not always automatic. For instance, polymer residues after transfer increase the measured height at the flake edge. A comparison of the height before and after transfer can thus serve as a direct measure of the thickness of the polymer residue.[115] Furthermore, the root mean square (rms) roughness can help to determine the quality of a transferred film. A lower rms value corresponds to a smoother surface, which is required for constructing heterostructures with pristine interfaces.

Strictly speaking, STM provides a map of the local density of states of a material, which most of the time can be interpreted as topographical information (though its interpretation is usually less direct than AFM). As the tunneling current is exponentially dependent upon the tip-sample distance, STM affords a very high resolution, making it a very precise surface imaging technique. A related technique, known as scanning tunneling spectroscopy (STS), can also be used in the same set-up, providing information on the local electronic densities of state of the material. The combination of both STM and STS was used by Delač Marion et al. on transferred ML CVD-grown MoS$_2$ on Ir(111).[171] A corrugation of the MoS$_2$ film was observed in the STM image, indicating a rippling of the 2D layer in a manner similar to the rippling of graphene on SiO$_2$/Si (see **Figure 18**(a) and (b)). Using STS, the authors observed an electronic bandgap consistent with a quasi-freestanding MoS$_2$ film, yielding information on the coupling to the metal substrate. The STS spectra are shown in Figure 18(c). Furthermore, Kerelsky et al. found spatially growing metal-induced gap states (MIGS) at the junction of ML MoS$_2$ and graphite using STS, within 2 nm of the junction.[194] Figure 18(d) and (e) show an STS color map and individual STS spectra taken at different distances from the MoS$_2$-graphite junction, respectively. This highlights the excellent resolution of STM/STS.

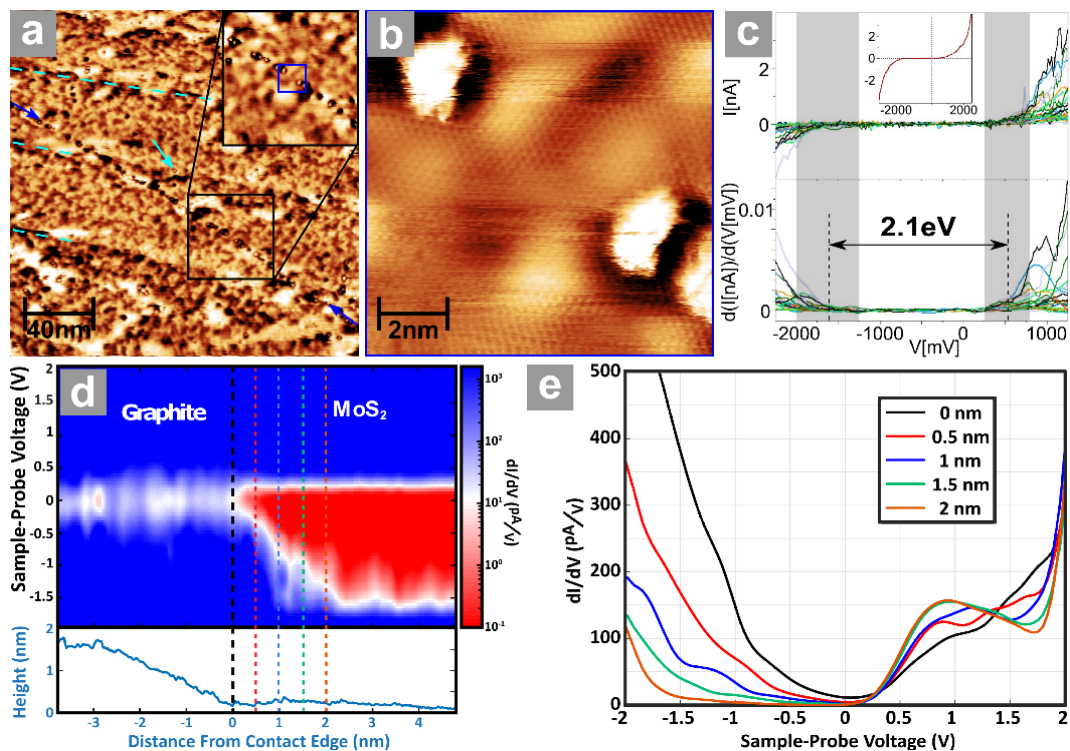

**Figure 18.** (a) Large area STM scan of ML MoS$_2$ on Ir(111). The pale blue dashed lines indicate the Ir steps underneath, with point defects specified by pale blue and dark blue arrows. (b) Smaller scale zoom image of inset in (a), showing two point defects. STM image showing the corrugation of the MoS$_2$ film. (c) STS spectra showing the dispersion of the gap edges. (d) STS map taken at the MoS$_2$/graphite junction, showing a gradual change in the LDOS within a short

distance of the contact edge. Individual STS spectra at various distances from the junction (shown as dashed lines in (d)), indicating a full gap opening only at several nanometers from the junction. (a-c) Reproduced with permission.[171] Copyright 2018, IOP Publishing. (d, e) Reproduced with permission.[194] Copyright 2017, American Chemical Society.

*3.2.4 Other characterization techniques*

In addition to the techniques mentioned above, other characterization methods are available to investigate the quality of transferred TMDs. This is motivated by the need for a versatile toolkit with which to characterize transferred TMDs. Here we discuss a few which provide valuable information concomitant with those offered by Raman, PL and SPM.

Firstly, information on structural defects, such as point defects, grain boundaries and dislocations, can be achieved with atomic resolution using transmission electron microscopy (TEM), in addition to chemical identification.[195,196] However, the technique is expensive and requires a special substrate. Additionally, features such as grain boundaries can also be imaged nondestructively using AFM.[196] Whilst surface roughness and domains can be observed locally using SPM, dark-field microscopy (DFM) can provide similar information on a larger scale.[16,65] Bubbles and wrinkles between stacked TMD interfaces scatter light, allowing for optical characterization. To this end, Kang et al. used DFM to characterize the density of scatterers between transferred 2L-$MoS_2$ interfaces, for two different transfer methods.[16] The vacuum stacking procedure (see Section 2.1.1) yielded > 99% reduction in the number of scatterers, compared to a dry transfer procedure. Notably, both films showed clear surfaces under standard bright-field (BF) imaging, indicating the advantage of DFM. In addition to DFM, non-linear optical microscopy techniques present a number of advantages. For instance, dark-field second harmonic generation (DF-SHG) can provide rapid and efficient mapping of 1D defects in ML-TMDs, which are not achievable with either BF or DF. This was done by Carvalho et al. to produce large scale spatial mapping of 1D defects in CVD-grown $MoSe_2$, $MoS_2$ and $WS_2$.[197] Moreover, the metallic character of the 1D grain boundaries in $MoSe_2$ could be determined by an enhancement of the SHG signal below the direct bandgap. Despite these notable advantages, such SHG methods are not effective at resolving grain boundaries when the crystal axis rotation between neighboring domains is small. Moreover, SHG is only observed in odd-layer films due to the breaking of inversion symmetry. To this end, third harmonic generation (THG) can allow for film characterization for both odd and even layer number.[198–200] Strikingly, the THG from ML-$MoS_2$ was observed to be 30 times stronger than the SHG, and about three times higher than for graphene.[199]

## 4. Conclusion and outlook

Significant progress has been made in the transfer of CVD-grown TMDs onto a variety of substrates over the last decade. Since the early-stage use of polymer supporting layers such as PMMA to transfer CVD-grown graphene[59], the method was successfully adopted to CVD-grown $MoS_2$[61] and thereafter extended. There now exist a range of supporting layers beyond PMMA that can be used to transfer large area films of 2D TMDs, from flexible low surface energy polymers such as PDMS[19,76,79], to more environmentally friendly natural polymers such as CA[99] and even to rigid non-polymer supports such as thin Cu films[44,105]. Moreover, there now exist methods which forgo the use of any support, relying on surface assisted delamination in an aqueous solution[92]. Each of these methods comes with its own advantages and drawbacks, and should be carefully considered for its usage in future applications. It is important to point out that such transfer techniques are still in their infancy in light of the issues discussed in Section 2. In order to meet the standards required for technological applications, the process related drawbacks of each transfer method (such as residues, cracks or wrinkles) must be addressed, as outlined in Section 3.

The extent to which 2D TMDs and their heterostructures can be integrated into existing silicon-based devices will depend largely on their compatibility with existing complementary metal-oxide-semiconductor (CMOS) fabrication processes. FETs, perhaps the most important electronic devices, are Si-based. It is therefore preferable to augment Si-CMOS with 2D transistors and heterostructures, rather than replace them. Existing CVD growth processes for TMDs require temperatures in the range of 600-900 °C[14], which limits their direct growth onto CMOS-ready substrates. Although some techniques to lower the growth temperature exist[201], the quality of the resulting material remains to be fully addressed. This leaves open the possibility of using modified transfer techniques for wafer-scale integration of 2D TMD films with CMOS-ready target substrates. In order for such transfer techniques, which are non-standard in microelectronic fabrication methods, to be successful, scalable methods must be developed. These methods should be automated and result in clean interfaces. A promising technique in this respect is that of vacuum stacking, as proposed by Kang et al. and discussed in Section 2.1.1.[16] This method made use of a direct transfer procedure, circumventing the problems encountered in removing the supporting polymer film.

Furthermore, methods to assess the quality of the 2D TMD films, at various stages of the microelectronic fabrication process, are required. A number of methods were outlined in Section 3.2, each providing process specific information. However, not all methods will be transferable to an industrial context. Intrusive characterization techniques such as TEM are useful in a laboratory setting, but are not industrially feasible as it results in electron-beam damage to the sample. Therefore, less destructive methods are preferred. Here optical characterization techniques can be extremely valuable. They require minimal sample preparation, and can provide fast two or three-dimensional wafer mapping in addition to high spatial resolution (on the micrometer scale).[202] Information related to coverage and layer numbers can also be obtained.[203] PL and Raman can also provide relevant structural and electronic information with respect to film quality, as can more advanced optical techniques such as second harmonic generation[197] (see Section 3.2). These methods can be adapted to a cleanroom setting for *in-situ* monitoring of the 2D film fabrication process.

The enormous potential of CVD-grown TMDs, both in terms of their unique intrinsic material properties as well as scalability afforded by the growth process, is capable of being realized in future device applications. However, a number of significant challenges remain to be overcome. As long as direct growth methods using suitable substrates remains out of reach, transfer techniques will remain an important means by which 2D TMDs can be integrated into existing device architectures. Moreover, for certain applications that require flexible substrates (such as wearable electronics), TMD transfer will likely be the main fabrication route.

**Acknowledgements**

Adam J. Watson and Wenbo Lu contributed equally to this work.

This work was supported by the Dutch Science Foundation (NWO) by the research grants NWO Vici (680-47-633) and NWO Start-Up (STU.019.014), the Zernike Institute for Advanced Materials, the European Union's Horizon 2020 research and innovation programme under grant agreement No 785219 (Graphene Flagship Core 3), and the Chinese Scholarship Council CSC.

**Conflict of Interest**

The authors declare no conflict of interest.

**Table of Contents**

Recent progress in the transfer of wafer-scale 2D semiconductors is comprehensively reviewed. Focus is given to chemical vapor deposition (CVD) grown group-VI transition metal dichalcogenides (TMDs). Relative advantages and disadvantages of each transfer method, as well as characterization techniques to evaluate the transferred film quality, are discussed in detail. Industrial feasibility is proposed and presented.

*Adam J. Watson, Wenbo Lu, Marcos H. D. Guimarães\* and Meike Stöhr\**

**Transfer of Large-Scale Two-Dimensional Semiconductors: Challenges and Developments**

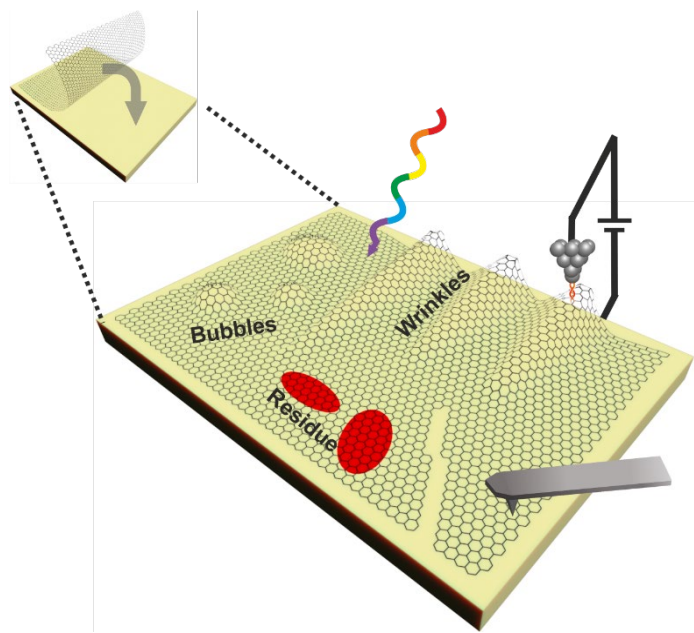